\documentclass[review]{elsarticle}

\usepackage{hyperref}
\usepackage{subfig}

\journal{Computer Methods and Programs in Biomedicine}

\usepackage{lipsum}
\makeatletter
\def\ps@pprintTitle{%
 \let\@oddhead\@empty
 \let\@evenhead\@empty
 \def\@oddfoot{}%
 \let\@evenfoot\@oddfoot}
\makeatother

\usepackage{tabularx}
\usepackage{multirow}
\usepackage{booktabs}
\usepackage{blindtext}
\usepackage{graphicx}
\usepackage{booktabs}
\usepackage{tabulary} 

\usepackage{svg}

\usepackage{tikz}
\usetikzlibrary{spy}

\begin{document}
    
    \begin{frontmatter}
    
        \title{Going Deeper through the Gleason Scoring Scale: An Automatic end-to-end System for Histology Prostate Grading and Cribriform Pattern Detection}
        
        \cortext[]{This work was supported by the Spanish Ministry of Economy and Competitiveness through project DPI2016-77869. The Titan V used for this research was donated by the NVIDIA Corporation.\\
         \copyright 2020. This manuscript version is made available under the CC-BY-NC-ND 4.0 license http://creativecommons.org/licenses/by-nc-nd/4.0/.}
        
        \author[aff1]{Julio Silva-Rodr\'iguez}
        \ead{jjsilva@upv.es}
        
        \author[aff2]{Adri\'an Colomer}
        \ead{adcogra@ui3b.upv.es}
        
        \author[aff3]{Mar\'ia A. Sales}
        \ead{salesman@gva.es}

        \author[aff4]{Rafael Molina}
        \ead{rms@decsai.ugr.es}
        
        \author[aff2]{Valery Naranjo}
        \ead{vnaranjo@dcom.upv.es}
        
        \address[aff1]{Institute of Transport and Territory, \textit{Universitat Polit\`ecnica de Val\`encia}, Valencia, Spain}
        
        \address[aff2]{Institute of Research and Innovation in Bioengineering, \textit{Universitat Polit\`ecnica de Val\`encia}, Valencia, Spain}
        
        \address[aff3]{Anatomical Pathology Service, University Clinical Hospital of Valencia, Valencia, Spain}
        
        \address[aff4]{Department of Computer Science and Artificial Intelligence, University of Granada, Granada, Spain}
        
        \begin{abstract}
            \textit{Background and Objective:}\\
Prostate cancer is one of the most common diseases affecting men worldwide. The Gleason scoring system is the primary diagnostic and prognostic tool for prostate cancer. Furthermore, recent reports indicate that the presence of patterns of the Gleason scale such as the cribriform pattern may also correlate with a worse prognosis compared to other patterns belonging to the Gleason grade $4$. Current clinical guidelines have indicated the convenience of highlight its presence during the analysis of biopsies. All these requirements suppose a great workload for the pathologist during the analysis of each sample, which is based on the pathologist's visual analysis of the morphology and organisation of the glands in the tissue, a time-consuming and subjective task. 

In recent years, with the development of digitisation devices, the use of computer vision techniques for the analysis of biopsies has increased. However, to the best of the authors' knowledge, the development of algorithms to automatically detect individual cribriform patterns belonging to Gleason grade $4$ has not yet been studied in the literature. The objective of the work presented in this paper is to develop a deep-learning-based system able to support pathologists in the daily analysis of prostate biopsies. This analysis must include the Gleason grading of local structures, the detection of cribriform patterns, and the Gleason scoring of the whole biopsy.

\textit{Methods:}\\
The methodological core of this work is a patch-wise predictive model based on convolutional neural networks able to determine the presence of cancerous patterns based on the Gleason grading system. In particular, we train from scratch a simple self-design architecture with three filters and a top model with global-max pooling. The cribriform pattern is detected by retraining the set of filters of the last convolutional layer in the network. Subsequently, a biopsy-level prediction map is reconstructed by bi-linear interpolation of the patch-level prediction of the Gleason grades. In addition, from the reconstructed prediction map, we compute the percentage of each Gleason grade in the tissue to feed a multi-layer perceptron which provides a biopsy-level score.

\textit{Results:}\\
In our SICAPv2 database, composed of $182$ annotated whole slide images, we obtained a Cohen's quadratic kappa of $0.77$ in the test set for the patch-level Gleason grading with the proposed architecture trained from scratch. Our results outperform previous ones reported in the literature. Furthermore, this model reaches the level of fine-tuned state-of-the-art architectures in a patient-based four groups cross validation. In the cribriform pattern detection task, we obtained an area under ROC curve of $0.82$. Regarding the biopsy Gleason scoring, we achieved a quadratic Cohen's Kappa of $0.81$ in the test subset.

\textit{Conclusions:}\\
Shallow CNN architectures trained from scratch outperform current state-of-the-art methods for Gleason grades classification. Our proposed model is capable of characterising the different Gleason grades in prostate tissue by extracting low-level features through three basic blocks (i.e. convolutional layer + max pooling). The use of global-max pooling to reduce each activation map has shown to be a key factor for reducing complexity in the model and avoiding overfitting. Regarding the Gleason scoring of biopsies, a multi-layer perceptron has shown to better model the decision-making of pathologists than previous simpler models used in the literature.

        \end{abstract}
        
        \begin{keyword}
        prostate cancer, Gleason, cribriform, Whole Side Images, convolutional neural networks, deep learning.
        \end{keyword}
        
    \end{frontmatter}
    

\section{Introduction}

\begin{figure*}[htb]
    \centering
    
      \subfloat[\label{fig1a}]{\includegraphics[width=.22\linewidth]{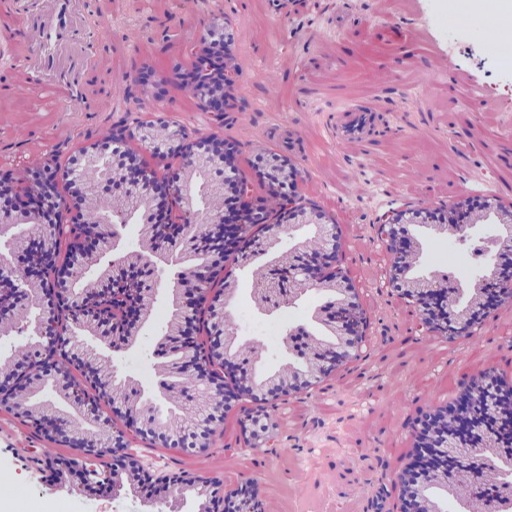}}
      \hspace*{\fill}
      \subfloat[\label{fig1b}]{\includegraphics[width=.22\linewidth]{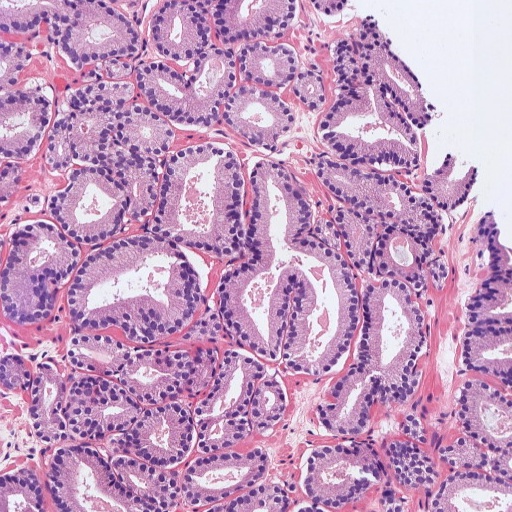}}
      \hspace*{\fill}
      \subfloat[\label{fig1c}]{\includegraphics[width=.22\linewidth]{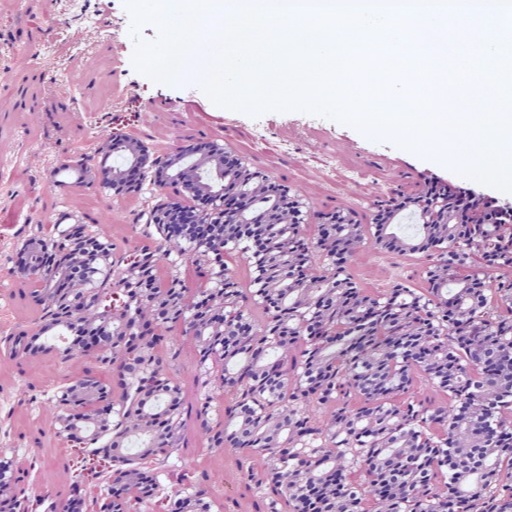}}
      \hspace*{\fill}
      \subfloat[\label{fig1d}]{\includegraphics[width=.22\linewidth]{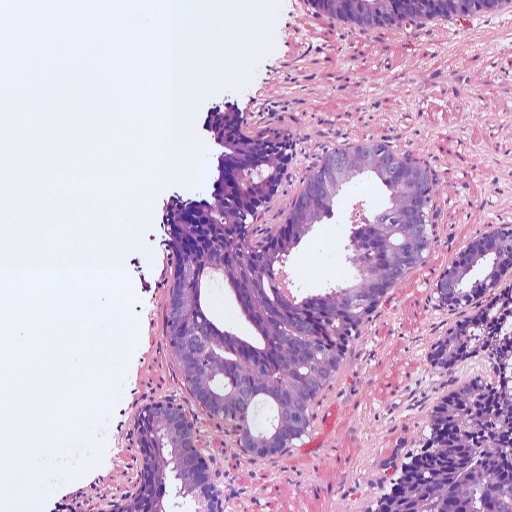}}
      \hspace*{\fill}
      
      \subfloat[\label{fig1e}]{\includegraphics[width=.22\linewidth]{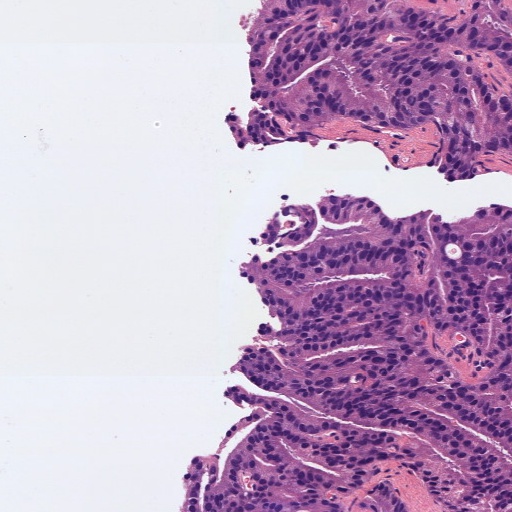}}
      \hspace*{\fill}
      \subfloat[\label{fig1f}]{\includegraphics[width=.22\linewidth]{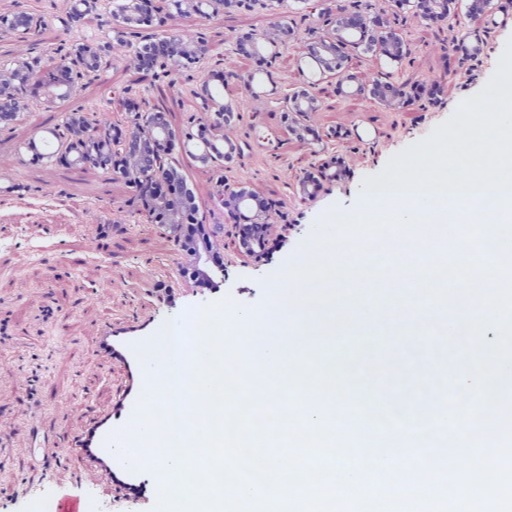}}
      \hspace*{\fill}
      \subfloat[\label{fig1g}]{\includegraphics[width=.22\linewidth]{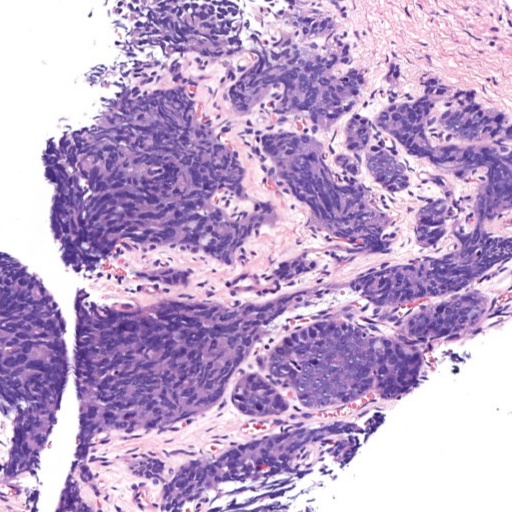}}
      \hspace*{\fill}
      \subfloat[\label{fig1h}]{\includegraphics[width=.22\linewidth]{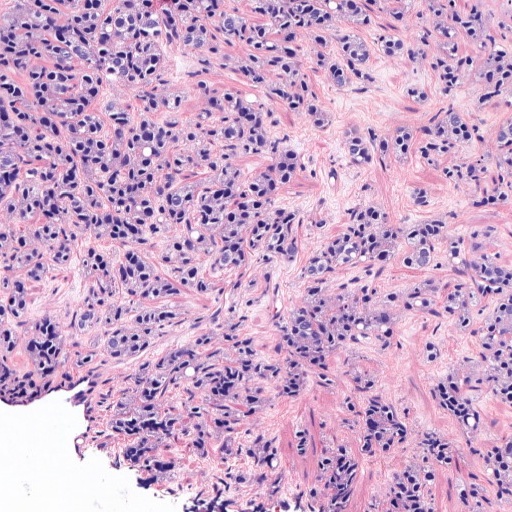}}
      \hspace*{\fill}
      
    \caption{Patches of H\&E histology samples presenting different Gleason patterns. (a): Non-cancerous well-differentiated glands; (b): Gleason grade $3$ containing atrophic dense patterns; (c): Gleason grade $4$ containing large fused glandular patterns; (d): Gleason grade $4$ containing cribriform patterns; (e): Gleason grade 4 containing papillary structures; (f): Gleason grade 4 containing individual poorly-formed glands; (g): Gleason grade 5 including nests of cells without lumen formation; (g): Gleason grade 5 containing files of isolated cells.}
    
    \label{fig1}
\end{figure*}

Worldwide, prostate cancer (PCa) is the second most common cancer in men, with $1.3$ million new patients in $2018$ \cite{wcrf}. According to the World Health Organisation, the yearly number of new cases will increase by more than $40\%$ in this decade \cite{who}. The main tool to diagnose PCa, once clinical explorations or blood test suggest its presence, is the prostate biopsy. Small portions of the tissue are extracted with a needle, laminated, stained with Hematoxylin and Eosin (H\&E) and finally stored in crystal. Then, the sample is analysed under the microscope by the pathologist, determining the presence and grade of cancerous patterns depending on the morphology and organisation of the glands, nuclei and lumen using the Gleason grading system \cite{gleason}. In this system, different cancer patterns in the tissue are grouped in different grades according to the prognosis of the cancer. In particular, for two-dimensional tissue slides, the Gleason grades (GG) range from $3$ to $5$, correlating inversely with the degree of gland differentiation of the tissue. The Gleason grade $3$ (GG3) includes atrophic well differentiated and dense glandular regions. The GG4 contains cribriform, ill-formed, large-fused and  papillary glandular patterns. Finally, GG5 includes isolated cells or file of cells, nests of cells without lumina formation and pseudo-roseting patterns. Examples of   patterns belonging to different grades are presented in Figure \ref{fig1}.     

Pathologists classify by visual inspection the tissue regions, detecting the presence of one or more Gleason patterns and, finally, diagnose the combined Gleason score according to the most prominent grades (e.g. the combined grade $5+4=9$ would be assigned to a sample in which the main cancerous Gleason grade is $5$ followed by the grade $4$). Therefore, the combined Gleason score ranges from $6$ to $10$, and it is assigned to the whole biopsy. This score is currently the best marker of prostate cancer prognosis and it defines the treatment to apply \cite{Gordetsky2016GradingImplications}. However, the Gleason scoring of histological prostate biopsies is a high time-consuming and repetitive task, which has intra and inter pathologist variability. Moreover, after the last International Society of Urological Pathology (ISUP) Consensus Conference in $2014$ \cite{Epstein2016TheSystem}, new guidelines have been included that increase the pathologists' workload. In particular, it is recommended to also report the percentage of Gleason grade $4$ in the sample, mainly for regions scored as $3+4=7$, where a higher percentage of Gleason grade $4$ indicates the convenience of an earlier treatment \cite{Sharma2018PercentCancer}, and the presence of cribriform glandular patterns, which indicate worse prognosis than the presence of other Gleason grade $4$ patterns \cite{Hassan2018ClinicalCancer,vanderKwast2018OnCancer}. Computer-Aided Diagnosis systems (CAD) support the work of pathologists and increase the objectivity in the this process. These are based on the digitisation of the histological crystals, obtaining whole slide images (WSIs) and developing computer vision algorithms to detect the cancerous regions inside the biopsy (or WSI).

\bigskip
\bigskip

Computer vision algorithms have been widely used to analyse histological PCa images. This section summarises the works previously presented in the CADs literature for prostate cancer detection, classifying them according to three factors: the kind of images included in the analysed database, the objectives addressed by CAD systems, and the techniques proposed to achieve them.

Regarding the images, mainly three types of histological images have been used: WSIs, prostactetomies and Tissue Micro Arrays (TMAs). TMAs are clusters of representative tumor areas extracted manually by pathologists \cite{Remotti2013TissueUse}. TMAs are used for testing new techniques in a large number of different tumour samples. One of the main limitations of TMAs lies in the small amount of tissue that can be included in each samples, which may not be representative of the whole tumor region in epithelial tumors with heterogeneous patterns \cite{KHOUjA2010LimitationsAnalysis}. This is the case of prostate cancer, which has different patterns for each Gleason grade, as previously mentioned. Non-cancerous patterns that could confuse CAD systems, as the inflamed tissue or benign multi-nucleation, could be lost using TMAs. Thus, the strategy based on TMA analysis is not used in clinical practice \cite{Voduc2009TissueOncology} and it is more convenient to develop CAD systems based on raw WSI analysis. A model trained using large databases of WSIs could be used for both WSIs and prostactetomies. The works in \cite{Doyle2007AutomatedFeatures,Gertych2015MachineProstatectomies,JimenezdelToro2017ConvolutionalScore,Ren2017ComputerSystem,Ing2018SemanticNetworks,Esteban2019AProcesses,Lucas2019DeepBiopsies} follow the strategy of WSI analysis, while in \cite{Arvaniti2018AutomatedLearning,Nir2018AutomaticExperts,Nir2019ComparisonImages} the authors use TMAs to develop the CAD models.

 With regard to the objectives to be addressed, some works focus just on the detection of prostate cancer against non-cancerous tissue \cite{Gertych2015MachineProstatectomies,Esteban2019AProcesses} or on the first-stage prostate cancer detection \cite{Garcia2019First-stageLearning}. A full analysis of Gleason grades from $3$ to $5$ is usually limited by the size of the collected database, and the low prevalence of Gleason grade $5$. Due to that, numerous researchers classify differentiating among non-cancerous samples, low grade (Gleason grade $3$), and high Grade (Gleason grade $\geq 4$) \cite{Lucas2019DeepBiopsies,Ma2018GeneratingRetrieval,Li2019PathImages} or among non-cancerous, Gleason grade $3$, and Gleason grade $4$ \cite{Doyle2007AutomatedFeatures,Ren2017ComputerSystem}. The most recent works tried to predict the full Gleason grading  (Benign - Grade $3$ - Grade $4$ - Grade $5$) in \cite{Arvaniti2018AutomatedLearning,Nir2018AutomaticExperts,Nir2019ComparisonImages} but only using TMAs cores. To the best of the authors's knowledge, works analysing deeper the Gleason grades, this is, focusing on the automatic detection of individual patterns of a Gleason grade (i.e. cribriform pattern, which belongs to the Gleason grade $4$ group) do not exist. This work represents an attempt in this direction.

\begin{figure}[htb]
    \centering
    \includegraphics[width=1\textwidth]{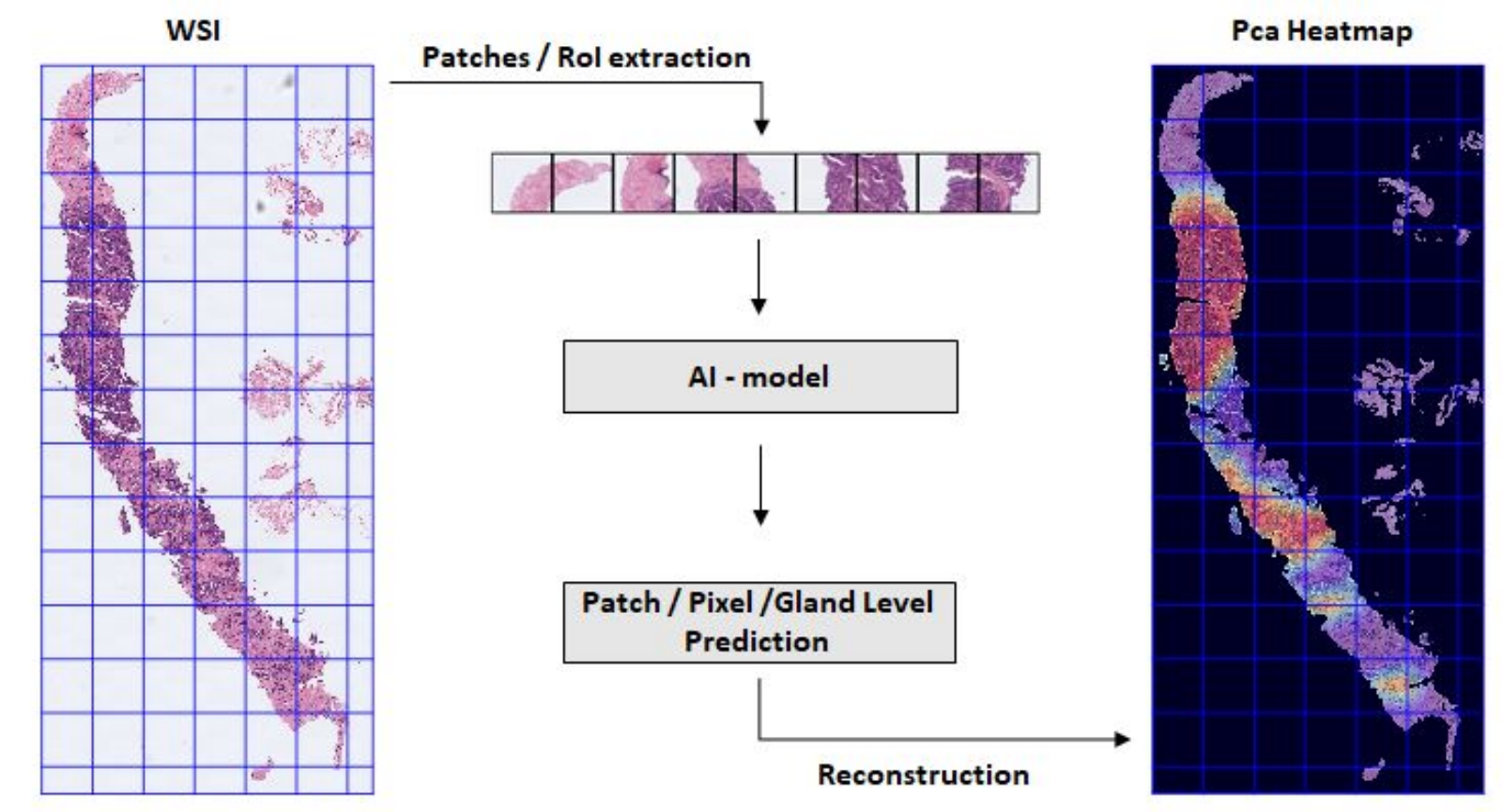}
    \caption{General workflow for high resolution histology slides processing.}
    \label{fig3}
\end{figure}

Finally, concerning the techniques used to deal with the different mentioned objectives, the most common approach to analysed both is to perform a patch-based strategy (see Figure \ref{fig3}). The motivation for using this strategy is the large size of both TMAs and, especially WSIs, together with hardware limitations. 

Below, we will focus only on the description of the different techniques used, until now, for the patch-level Gleason grading. In the literature we can find approaches based on classic machine learning techniques with a hand-crafted feature extraction and deep learning algorithms (automatic feature extraction) by means of convolutional neural networks (CNN). In Nir et al. ($2018$) \cite{Nir2018AutomaticExperts} a comparison between both approaches is carried out with a database of $333$ cores of TMAs. Glands and nuclei are segmented to obtain features related to their size, intensity distributions and number of elements in each patch at different resolutions. Those are combined with full patch-level features related to the colour distribution and SURF descriptors to fit different machine learning models as linear discriminant analysis, linear regression, support vector machines, and random forests. Those models are compared with a U-Net CNN. The best result reported is a Cohen's quadratic kappa ($\kappa$) overall agreement measure of $0.51$ obtained by the linear regression model. Nevertheless, in a later publication by Nir et al. ($2019$) \cite{Nir2019ComparisonImages} a $\kappa$ of $0.60$ was obtained by fine-tuning the CNN architecture MobileNet. In Arvaniti et al. ($2018$) \cite{Arvaniti2018AutomatedLearning} a larger database is used, with $886$ cores. The patch-level grading is addressed through fine-tuning different CNN architectures such as VGG16, InceptionV3, ResNet50, DenseNet121, and MobileNet. The best results are reported with the last one, achieving a $\kappa$ of $0.67$ in the training set and $0.55$ in the test one.

Regarding the classification of the Gleason score for the whole biopsy (whole slide image), only a few works have addressed it, and only using TMAs. The common strategy used is to obtain the percentage of each grade in the analysed image and to assign the first and second components above a threshold as primary and secondary grades respectively. In Arvaniti et al. ($2018$) \cite{Arvaniti2018AutomatedLearning} the full Gleason scoring, using TMAs,  is addressed, archiving $\kappa$ of $0.75$. Unfortunately, this simple model did not perform for extreme cases, for example $5+5=10$. In this case, a precision of $0.10$ is reported in this work. In addition, the primary and secondary grades are not just related to the proportion of the different grades in the tissue, but also to the severity of each grade (e.g. GG5 could be diagnosed as secondary grade even having less proportion than GG4 or GG3 in the tissue).
\bigskip
\bigskip

The objective of this work is to develop an automatic Computer-Aided Diagnosis system working on WSIs and able to support pathologists in the analysis of the biopsy during the diagnosis process. The tasks of this analysis, to be included in the pathologists' report, are:

\begin{itemize}

    \item Detection of the cancerous regions in the tissue according to the Gleason grading system.
    
    \item Detection of cribriform patterns.
    
    \item Calculation of the percentage of each Gleason grade in the biopsy.
    
    \item Gleason scoring of the whole biopsy, taking into account not only the grade proportion but also its severity.
    
\end{itemize}

\begin{figure}[htb]
    \centering
    \includegraphics[width=1\textwidth]{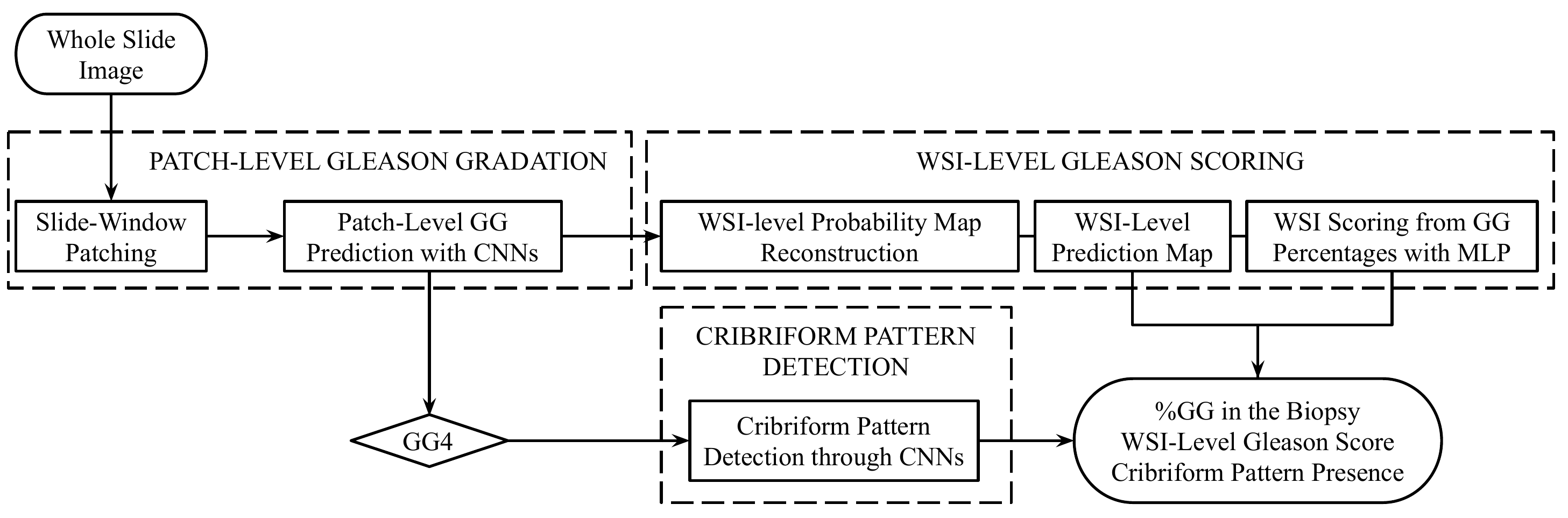}
    \caption{Flowchart in which the different blocks of our system are presented. Taking as input a prostate whole slide image (WSI), the system performs a patch-level Gleason grade prediction through convolutional neural networks. If one patch is classified as Gleason grade $4$ (GG4), a cribriform pattern detection is carried out by fine-tuning the model of the previous stage. Finally, the regions in the WSI are reconstructed and a pixel-level Gleason grade assignement is carried out. The WSI-level Gleason scoring is performed with a multi-layer perceptron taking as input the percentage of the Gleason grades in that region.}
    \label{work_flowchart}
\end{figure}

This work is developed using our collected database SICAPv2, the largest public database of prostate biopsies with pixel-level annotations of Gleason grades, specifying the presence of cribriform patterns. In the following lines, we summarise the main contributions of this paper. The different blocks of our system are presented in Figure \ref{work_flowchart}. First, we develop a patch-level predictor of Gleason grades with a carefully-designed CNN architecture trained from scratch. This architecture is based on three convolutional blocks and global-max pooling after the last block. With this model, we outperform, for the first time in the literature, the fine-tunning well-known state of the art architectures. Then, we discuss the model interpretability by means of the Class Activation Maps (CAMs) technique. Once the patches are classified, the trained architecture is fine-tuned to detect the presence of cribriform glandular structures for those images with Gleason grade $4$. To the best of the authors' knowledge, no study has addressed this clinical need previously. Then, the WSIs are reconstructed in probability maps and the class (i.e. non cancerous, Gleason grade $3$, $4$ or $5$) with the highest probability is assigned to each pixel. Once the percentages of each Gleason grade in the WSI are obtained, we developed a model, based on a multi-layer perceptron architecture, to predict the combined Gleason score to the whole biopsy. The obtained results show the good performance of this model which outperforms the previous state-of-the-art methods.


The paper is organised as follows, in Section \ref{materials} we introduce the database used in this work, SICAPv2, a large set of prostate whole slide images with pixel-level annotations of the Gleason grades and WSI-level annotations of the Gleason scores assigned by expert pathologists. In Section \ref{methods} we describe the methodological details of our proposed CAD system, based on CNNs able to predict the Gleason grade and presence of cribriform pattern in local patches of the WSIs. From those local predictions, in this section we also detail the process of predicting the WSI-level Gleason score. In Section \ref{experiments} we describe the performed experiments in order to validate our models. In particular, Section \ref{chap:patchLevel} describes the experiments related to the patch-level Gleason grading, Section \ref{cribriform} the detection of cribriform patterns and in Section \ref{region} we present our results related to the biopsy-level Gleason scoring. Finally, Section \ref{conclusions} summarises the conclusions extracted with the carried out experiments.

\section{Materials: SICAP database}
\label{materials}

 The database presented in this paper, SICAPv2, is, to the best of the authors's knowledge, the largest public collection of prostate H\&E biopsies with local-level annotations of Gleason grades. SICAPv2 is an extension the database introduced in \cite{Esteban2019AProcesses} and will be publicy available after the publication of this paper.
 
 After analysing the literature, four main prostate cancer tissue image databases were found. The largest database with prostate biopsies was released by The Cancer Genome Atlas project\footnote{\url{https://portal.gdc.cancer.gov/}} \cite{Weintein2013TheProject} with up to $720$ prostate biopsy slides. Nevertheless, the lack of annotations at both the local and biopsy levels of the Gleason grades restricts the use of these data. The database shared by Arvaniti et al. \cite{Arvaniti2018AutomatedLearning} includes pixel-level annotations of Gleason patterns from $886$ small regions of slides (cores of TMAs). Unfortunately, as discussed earlier, those cores do not represent the heterogeneous patterns of local structures of prostate cancer and benign lesions, so they lack clinical relevance for the slide-level Gleason score diagnosis. Similar limitations are found in the recent database from the challenge Gleason19 in the MICCAI $2019$ conference\footnote{\url{https://gleason2019.grand-challenge.org/Home/}}, with $331$ cores annotated by different pathologists, and the dataset used in \cite{Ing2018SemanticNetworks}, composed by $625$ isolated patches. Although those databases contribute to the validation of different algorithms, the lack of large databases with clinical reference of heterogeneous patterns has been a limiting factor for the scientific community to develop deep-learning-based methods which demand a large amount of data. One of the contributions of this work is the publication of a large database of WSIs containing biopsy-level labels (i.e. Gleason scores for each biopsy) and pixel-level Gleason grades annotations, in which for the first time, the presence of cribriform glandular regions is indicated.

\begin{figure}[htb]
      \centering
      \includegraphics[width=.5\linewidth]{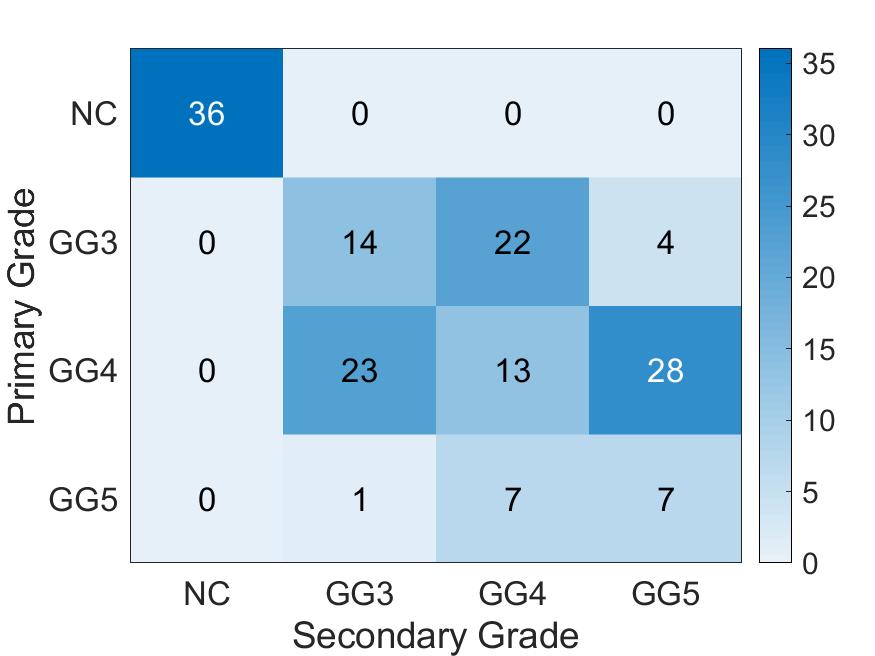}
    \caption{Description of the Gleason scores in the SICAPv2 database. Co-occurrence matrix of primary and secondary Gleason grades in each biopsy. NC: non cancerous, GG3: Gleason grade 3, GG4: Gleason grade 4 and GG5: Gleason grade 5.}
    \label{fig:coocurrence}
\end{figure}

SICAPv2 database includes $155$ biopsies from $95$ different patients who signed the pertinent informed consent. The tissue samples where sliced, stained and digitised using the Ventana iScan Coreo scanner at $40x$ magnification obtaining WSIs. The slides were analysed by a group of expert urogenital pathologists at Hospital Cl\'inico of Valencia, and a combined Gleason score was assigned per biopsy. In cases where the grade was uncertain, the label was assigned by consensus of all expert pathologists to avoid inter-observer variability. The primary Gleason grade (GG) in each biopsy is distributed as follows: $36$ non-cancerous regions, $40$ samples with Gleason grade $3$, $64$ with Gleason grade $4$ and $15$ with Gleason grade $5$ (henceforth NC, GG3, GG4, and GG5 respectively). Regarding the combined scores, the co-occurrence matrix of primary and secondary grades is shown in Figure \ref{fig:coocurrence}.

The local cancerous patterns were annotated using an in-house software based on the OpenSeadragon libraries \cite{microdraws}, following the Gleason scale and indicating the presence of cribriform glandular structures. In order to process the large WSIs, these were down-sampled to $10x$ resolution and divided into patches of size $512^2$ and overlap of $50\%$ between them. Those values were previously optimised for the detection of cancerous patterns in \cite{Esteban2019AProcesses}. A mask of the presence of tissue in the patches was obtained by applying the Otsu threshold method. To develop the model able to predict the main Gleason grade, patches with less than $20\%$ of tissue were excluded. In addition, patches without cancerous patterns annotated by the pathologists belonging to cancerous biopsies where also discarded. After this procedure, the database contains $4417$ non-cancerous patches, $1635$ labelled as GG3, $3622$ as GG4, and $665$ as GG5. Note that if one patched contained more than one annotated grade, the majority grade was assigned as label. 763 GG4 patches also contain annotated cribriform glandular regions. A summary of the database description is presented in Table \ref{table:datasets}.

\begin{table}[htb]
\centering
\caption{SICAPv2 database description. Amount of whole slide images and their respective biopsy-level primary label (first row) and number of patches of each one of the Gleason categories (second row).}
\label{table:datasets}
\resizebox{\linewidth}{!}{
\begin{tabular}{lccccc}
\toprule
                    & \textbf{Non cancerous} & \textbf{Grade 3} & \textbf{Grade 4 (cribriform)} & \textbf{Grade 5} & \textbf{Total} \\
\midrule
\midrule
\textbf{\#WSIs}     & $37$                   & $60$             & $69$ $(36)$                          & $16$             & $182$          \\
\textbf{\#Patches}  & $4417$                 & $1636$           & $3622$ $(763)$                & $665$            & $10340$        \\
\bottomrule
\end{tabular}
}
\end{table}

The data collected by Arvaniti et al. \cite{Arvaniti2018AutomatedLearning} was also utilised to validate the models produced in our study. The cores were resized to match the resolution used in our models and patched to the dimensions used in our database. By this approach, each one of these cores is approximately equivalent to one of our patches. Thus, $115$ non-cancerous images, $274$ patches labelled as GG3, $210$ GG4, and $104$ GG5 were used to validate our work in an external database. Also, the patches shared by Gerytch et al. \cite{Gertych2015MachineProstatectomies} were used in our work for the validation of our proposed model. After normalisation of the images to match our methodology, $32$ non-cancerous images, $95$ patches labelled as GG3, $216$ GG4, and $70$ GG5 were obtained.

\section{Methods}
\label{methods}
\subsection{Patch-Level Gleason Grading}
\label{Patch-level}

The patch-level classification in the different Gleason grades is carried out by means of convolutional neural networks. We propose a self-designed base-model architecture (from now on called $FSConv$) which consists of a simple convolutional architecture with three convolutional layers and dimensional reduction operation employing max-pooling layers (Table \ref{gleasonNet1}).

\begin{table}[htb]
\centering
\caption{$FSConv$ architecture description. It consists of three blocks with convolutional filters, ReLU activation and max-pooling operation.}
\label{gleasonNet1}
\resizebox{\linewidth}{!}{
\begin{tabular}{lcccccc}
\toprule
\textbf{Layer Name} & \textbf{Filter Size} & \textbf{Stride} & \textbf{Activation} & \textbf{Output Shape} & \textbf{Connected to} \\
\midrule
\midrule
 $Input$            & $-$            & $-$ & $-$    & $(224, 224, 3)$   & $-$               \\
 $Conv_1$           & $(3, 3, 32)$   & $1$ & $ReLU$ & $(224, 224, 32)$  & $Input$           \\
 $Max-Pooling_1$    & $(2, 2)$       & $2$ & $-$    & $(112, 112, 32)$  & $Conv_1$          \\
 $Conv_2$           & $(3, 3, 124)$  & $1$ & $ReLU$ & $(112, 112, 124)$ & $Max-Pooling_1$   \\
 $Max-Pooling_2$    & $(2, 2)$       & $2$ & $-$    & $(56, 56, 124)$   & $Conv_2$          \\
 $Conv_3$           & $(3, 3, 512)$  & $1$ & $ReLU$ & $(56, 56, 512)$   & $Max-Pooling_2$   \\
 $Max-Pooling_3$    & $(2, 2)$       & $2$ & $-$    & $(28, 28, 512)$   & $Conv_3$          \\
\bottomrule
\end{tabular}
}
\end{table}

After the automatic feature extraction blocks (base model), we introduce as top model a global-max-pooling layer. To show the superior performance of this architecture, different configurations already applied in the literature to the same problem, have been also tested as top models and are described next.

\begin{figure}[htb]
    \centering
      \subfloat{\includegraphics[width=.32\linewidth]{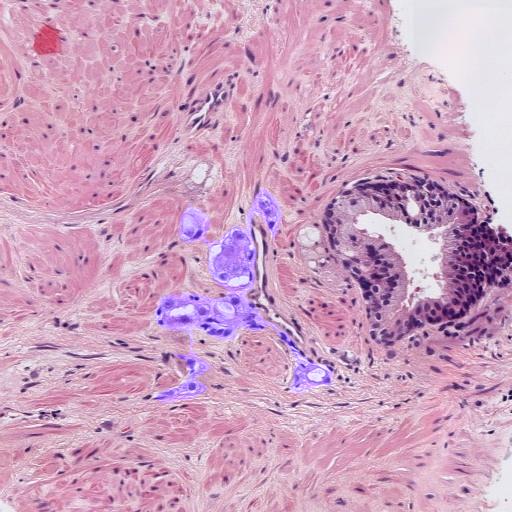}}
      \hspace*{\fill}
      \subfloat{\includegraphics[width=.32\linewidth]{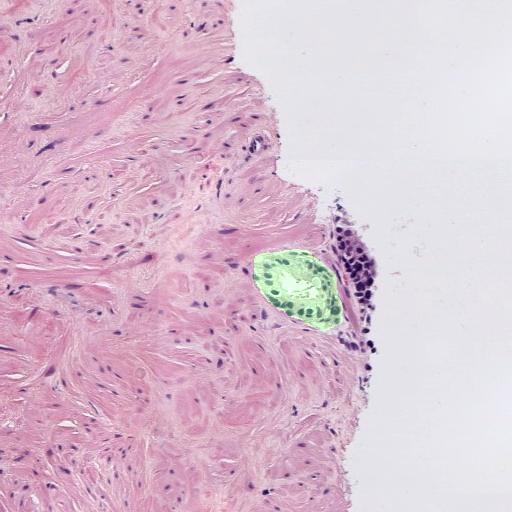}}
      \hspace*{\fill}
      \subfloat{\includegraphics[width=.32\linewidth]{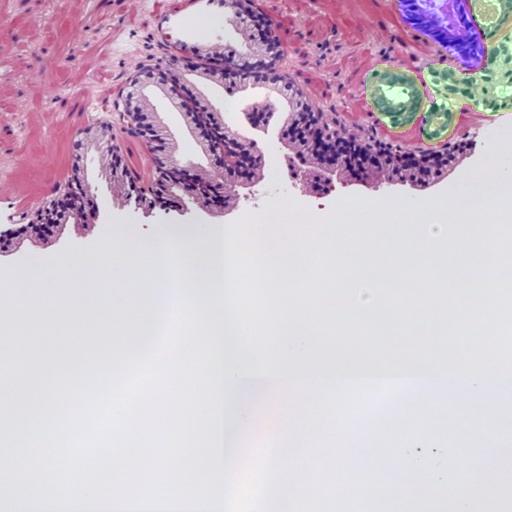}}
      \hspace*{\fill}
    \caption{Patches with small amount of cancerous tissue. Green: GG3, Blue: GG4.}
    \label{fig:amountTissue}
\end{figure}

One of the main approaches is the flattening of the activation volume resulting from the final convolutional block and the class prediction through consecutive fully-connected layers. In this case, overfitting is addressed by means of a random dropout of a percentage of the neurons in each training iteration. Nevertheless, these top-model architectures include a large number of parameters to optimise, increasing the complexity of the model, and they are sensitive to the location of the structures in the image. This problem is usually dealt with data augmentation techniques, applying, for example, random rotations and translations to the images. Other approaches propose the convenience of using global-average pooling on the last feature maps as regulariser to make the model translation-invariant and decrease its complexity \cite{Lin2014NetworkNetwork}. This technique is used in \cite{Arvaniti2018AutomatedLearning} for the prediction of prostate cancer Gleason degree with fine-tuned models. Due to the use of a patch-based strategy with sliding window, the location and amount of the cancerous structures in the image is not controlled. Thus, as shown in Figure \ref{fig:amountTissue}, some patches could have small portions of cancerous tissue. The global-average pooling layer takes into account the information in the whole activation map, and in those cases, the output of the filter that detects this pattern could be diminished. To make the models robust to the amount and location of cancerous tissue, we propose in this work the use of the global-max-pooling layer to play the role of the global-average pooling. All different configurations, fully-connected layer with ReLU activation and dropout regularisation (FC), global-average-pooling (GAP) and global-max-pooling (GMP) layers and their combinations are implemented and their performance is discussed in this work.

For comparison, together with the proposed architecture trained from scratch, we fine-tuned several well-known architectures: VGG19 \cite{Simonyan2014VeryRecognition}, ResNet-50 \cite{He2016DeepRecognition}, InceptionV3 \cite{Szegedy2016RethinkingVision} and MobileNetV2 \cite{Huang2017DenselyNetworks}. All of them were pre-trained in the Imagenet data set \cite{Deng2009ImageNet:Database}. For the feature extraction stage, the base model from those pre-trained models is extracted and partially retrained. This strategy is usually used to transfer the knowledge obtained in extracting features from a large database to specific domains where the amount of data is limited. Nevertheless, the patterns of the images used during the training are very different from the histology ones. To keep just the low-level features (contours, combination of basic colours, general shapes, etc.) from the pre-trained models, the weights of just the first convolutional blocks are frozen, while the rest are re-trained to adapt the model to the specific application. The layer from which the freezing strategy is applied is empirically optimised for each model, and is specified in the experimental part of the paper, in Section \ref{chap:patchLevel}.

The output layer for all the different configurations is composed of one neuron per class with soft-max activation function to obtain the final probability per class. In the training process, we use categorical cross-entropy as loss function, modified to deal with the class imbalance in the training set as follows:

\begin{equation}
\label{eq:loss}
L(\widehat{y},y) = - \frac{1}{C}\sum_{c=1}^{C}w_{c}(y_{c}log(\widehat{y}_{c}))
\end{equation}
\noindent where $y$ and $\widehat{y}$ contain the one-hot-encoded reference labels and predicted probabilities, respectively, of each class $c$ for a certain instance. $w_{c}= (C \times N) / N_{c}$ is the weight applied to each class, being $N$ the total number of images, $N_{c}$ the number of images belonging to class $c$ and $C$ the number of classes, $C=4$ in our case (non-cancerous, GG3, GG4 or GG5).

Stochastic Gradient Descend is applied as optimiser and the training procedure is performed using mini-batches. The values of learning rate and batch size are fixed empirically for each configuration and experiment, and they are specified in Section \ref{chap:patchLevel}. Data augmentation techniques are used on the training set applying random rotations and translations to the images.
\subsection{Cribriform Pattern Detection}
\label{chap:crib}

The detection of cribriform structures in GG4 patches is also carried out using convolutional neural networks. Due to the complexity of the task and the reduced number of samples, we address this problem by fine-tuning the model trained for the Gleason grades prediction. To take advantage of the specialised features extracted by the proposed architecture, the model is re-trained, optimising the layer from which the filter weights should be frozen to avoid over fitting. The top model used here is also proposed in the Gleason grading problem (global-max-pooling layer) followed by a last layer with one neuron and sigmoid activation function. The loss function used is the binary cross-entropy. Again, Stochastic Gradient Descent is used as optimiser applied on mini-batches and including data augmentation with random rotations, translations and brightness variations.
\subsection{Whole Slide Image Gleason Scoring}
\label{chap:WSIscoring}
To predict the Gleason score of the WSI, it is necessary to compute the tissue percentage of each Gleason grade present in the WSI. For that purpose, the first step is to apply the patch-level classification (section \ref{Patch-level}). Then, for each pixel, the predicted probabilities for each class is estimated by bilinearly interpolating the predicted probabilities of the closest patches in terms of euclidean distance to the center of the patches. Thus, a probability map per class (i.e NC, GG3, GG4, and GG5) is obtained per each WSI. Finally, the percentage of each Gleason grade is calculated after assigning each pixel the class, $c$, with the highest probability.





\begin{figure}[htb]
    \centering
      \includegraphics[width=.75\linewidth]{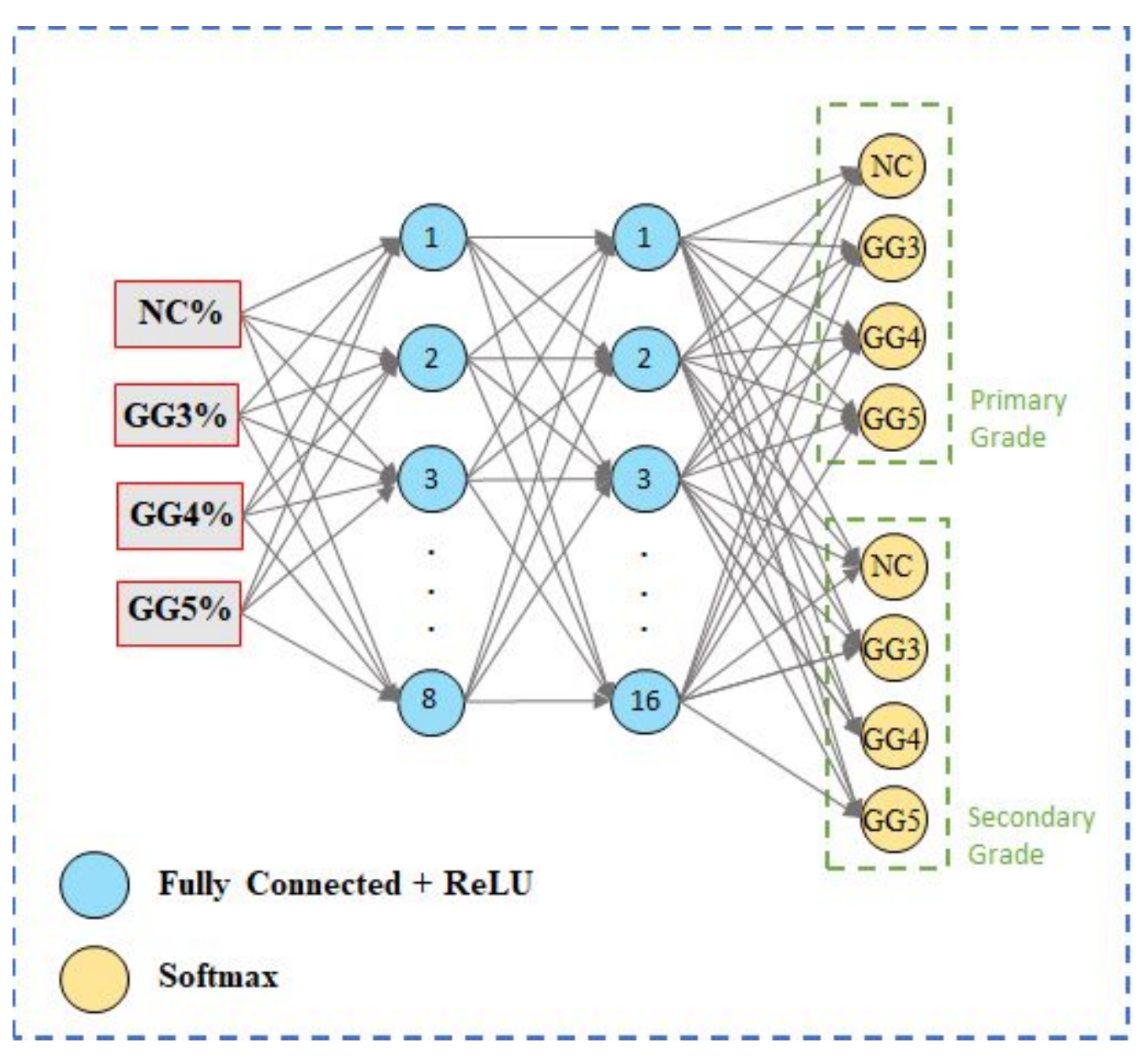}
    \caption{Proposed Multi-Layer Perceptron ($MLP$) for the whole slide image Gleason scoring. The model takes as input the percentage of each Gleason grade in the whole slide image, and is composed by a main branch with two fully-connected layers and two outputs. The intermediate layers consist of $8$ and $16$ neurons respectively and ReLU as activation function. The output layers present one neuron per target class and soft-max activation. NC: non cancerous, GG3: Gleason grade $3$, GG4: Gleason grade $4$, GG5: Gleason grade $5$.}
    \label{fig:mlp}
\end{figure}

The pathologist's decision making while assigning a Gleason score to a WSI takes into account both the percentage of each Gleason grade and the severity of each grade. To model this process, we propose to train a Multi-Layer Perceptron ($MLP$) to automatically predict the combined Gleason scoring of a biopsy, by means of a multi-class classification task. This task requires the prediction of both primary and secondary Gleason grades. To address it, MLP is selected as a suitable classifier, due to its flexibility to adapt the architecture to perform a multi-output classification. The proposed $MLP$ architecture consists of a branch with two outputs (see Figure \ref{fig:mlp}). The branch is composed of two fully-connected layers with $16$ and $8$ neurons respectively, and ReLU as activation function. The branch is then divided into two output layers: one for the primary Gleason grade and one for the secondary grade. These output layers are composed of four neurons each, one neuron per target class (i.e. NC, GG3, GG4 or GG5) and soft-max as activation function. The loss function used is the categorical cross-entropy.

\section{Experiments}
\label{experiments}

In this section we present the results of the different experiments carried out to show the performance of the proposed approach for the different classification tasks: patch-level classification, cribriform pattern detection and WSI scoring. In all cases, when possible, we also present a comparison with current state of the art methods and discuss the obtained results.

\subsection{Database Partitioning and Metrics}

In order to train the models and optimise the hyperparameters involved in this process, the database was divided following a cross-validation strategy. In particular, each patient was exclusively assigned to one fold with the aim of avoiding overestimation of the performance of the system \cite{Nir2019ComparisonImages} and ensuring its ability of generalisation. Thus, the database was divided into $5$ groups containing approximately $20\%$ of the patches each one. Notice that this process was carried out trying to guarantee the class balance character between sets. A summary of the resulting partition is presented in Table \ref{table:partition}.

\begin{table}[htb]
\centering
\caption{Database partition description: number of patients-patches for each grade in each validation fold ($4$-fold cross-validation) and test subset.}
\label{table:partition}
\begin{tabular}{lcccccc}
\toprule
 &  & \multicolumn{4}{c}{Patients - Patches} \\
\multicolumn{2}{c}{Group} & Non Cancerous & GG3 & GG4 (Cribriform) & GG5\\
\midrule
\midrule
\multirow{ 4}{*}{Cross-validation} & $1$ & $2$ - $685$  & $3$ - $625$ & $11$ - $979$ $(237)$ & $2$ - $198$     \\
                             & $2$ & $1$ - $717$  & $4$ - $346$ & $10$ - $950$ $(41)$  & $2$ - $153$     \\
                             & $3$ & $1$ - $644$  & $9$ - $361$ & $7$ - $670$ $(126)$  & $2$ - $118$     \\
                             & $4$ & $1$ - $1727$ & $8$ - $497$ & $9$ - $1042$ $(214)$ & $2$ - $247$     \\
Test                         &     & $4$ - $644$  & $6$ - $393$ & $9$ - $853$ $(145)$  & $2$ - $232$     \\
\bottomrule
\end{tabular}
\end{table}

Notice that four of the five sets were used to tune the hyper-parameters involved in the developed algorithms while the remaining partition was employed to test the final predictive system. For the evaluation of the patch-level Gleason grade prediction, a cross-validation strategy was used with the four validation cohorts, while for the WSI-level prediction of Gleason scores those sets were joined to apply a leave-one-out strategy per patient in training.

In order to objectively evaluate the performance of the trained models the following metrics were used: accuracy, F1-score, and Cohen's quadratic kappa statistic. The accuracy ($ACC$) is defined as the percentage of samples correctly classified. Nevertheless, this metric does not provide information about the performance of the model for each class. This information was quantified by utilising the F1-score ($F1S$), a combination of precision and sensitivity per class computed as follows:

\begin{equation}
\label{eq:f1s}
F1S_{c} = 2\times\frac{precision_{c}\times sensitivity_{c}}{precision_{c}+sensitivity_{c}}
\end{equation}
Cohe\noindent where $c$ indicates the predicted classes.

However, an automatic method should be less penalised when classifying a GG5 tissue as GG4 than as NC, even more so when taking into account the inter and intra-observer variability. In the literature, this fact is addressed using the Cohen's quadratic kappa ($\kappa$) metric \cite{Cohen1968WeightedCredit}. The metric $\kappa$ ranges from $-1$ to $1$, being directly proportional to the level of agreement between observers (-1 no agreement, 1 total agreement). Although there is not objective interpretation of which are the reasonable values for $\kappa$ in medical applications, recent proposals \cite{McHugh2012InterraterStatistic} define a moderate agreement if $\kappa$ is higher than $0.6$, while a strong agreement is stated when $\kappa$ is higher than $0.8$.

The patch-level Gleason grading models are evaluated using all the aforementioned figures of merit.

In order to evaluate the system for the detection of cribriform patterns, the area under the Receiver Operating Characteristic (ROC) curve ($AUC$) was used. In medical applications, a system is considered reliable if the $AUC$ value exceeds $0.80$ \cite{Swets1988MeasuringSystems}. The predicted labels are obtained by thresholding the scores (cribriform if the probability is above $0.5$), and then evaluated by means of $ACC$, sensitivity and specificity.

Regarding the evaluation of the WSI-level Gleason scoring, the Cohen's quadratic kappa was used.

\subsection{Patch-Level Gleason Grading}
\label{chap:patchLevel}
In the case of the patch-level Gleason grading model, in this section besides the obtained results using SICAPv2 database, we also discuss its performance in an external database. 

\subsubsection{FSConv Architecture Benchmarking}

After optimising the hyperparameters (learning rate, batch size, number of epochs, etc.), table \ref{res1} shows the obtained results in the validation sets for the proposed network $FSConv$ with different top models: fully-connected layers (FC), global-max pooling (GMP), global-average pooling (GAP), or a combination of them (GAP+FC or GMP+FC). Table \ref{res1} also presents the results for the best tested fine-tuned architectures, VGG19 and RestNet, using the same top models as $FSConv$. The optimum hyperparameters were: learning rate of $0.01$ for $FSConv$ and $0.0001$ for the finned-tunned networks, batch size of $32$ images and $200$ epochs in all cases. The base model of the fine-tuned networks were also optimised, being selected to freeze the first convolutional block for VGG19 and setting all layers as trainable for RestNet. Futhermore, Table \ref{models_parameters} presents a comparison in terms of storage space (in kilobytes, KB) and number of trainable parameters of each architecture.

\begin{table}[htb]
    \centering
    \caption{Results for patch-level Gleason grades prediction in the validation set. The performance of the different models ResNet, VGG19 and $FSConv$ are presented with the different configurations of top models. The metrics presented are the accuracy (ACC), the F1-Score (FS1), computed per class and its average, and the Cohen's quadratic kappa ($\kappa$). GMP: global-max pooling, GAP: global-average pooling and FC: fully-connected layers.}
    \label{res1}
    \resizebox{\linewidth}{!}{
    \begin{tabular}{|l|c|cccc|c|c|}
    \hline
    \multicolumn{1}{|c|}{\textbf{Experiment}} & \textbf{ACC} & \multicolumn{4}{c|}{\textbf{F1S}} & \textbf{Avg-F1S} & \textbf{$\kappa$}\\
     & & NC & GG3 & GG4 & GG5 & & \\
    \hline
    \hline
    
    VGG19+FC & $0.7218\pm0.0411$ & $\mathbf{0.8871\pm0.0178}$ & $0.6639\pm0.0509$ & $0.6041\pm0.1694$ & $0.5206\pm0.0996$ & $0.6689\pm0.0650$ & $0.7346\pm0.0324$               \\
    VGG19+GMP & $0.7213\pm0.0542$ & $0.8729\pm0.0207$ & $0.6480\pm0.0609$ & $0.6032\pm0.1673$ & $\mathbf{0.5450\pm0.0943}$ & $0.6673\pm0.0766$ & $0.7174\pm0.0641$               \\
    VGG19+GMP+FC & $0.7273\pm0.0424$ & $0.8860\pm0.0194$ & $0.6821\pm0.0633$ & $0.6093\pm0.1508$ & $0.5313\pm0.0820$ & $0.6772\pm0.0651$ & $\mathbf{0.7474\pm0.0648}$            \\
    VGG19+GAP & $0.7306\pm0.0460$ & $0.8814\pm0.0267$ & $0.6434\pm0.0961$ & $0.6530\pm0.1164$ & $0.5138\pm0.0847$ & $0.6729\pm0.0472$ & $0.7175\pm0.0730$                       \\
    VGG19+GAP+FC & $0.7246\pm0.0485$ & $0.8795\pm0.0130$ & $0.6905\pm0.0601$ & $0.6099\pm0.1542$ & $0.5216\pm0.1185$ & $0.6754\pm0.0724$ & $0.7179\pm0.0623$                    \\
    
    \hline
    
    ResNet+FC & $0.6952\pm0.0316$ & $0.8383\pm0.0151$ & $0.6670\pm0.0753$ & $0.5726\pm0.1271$ & $0.4845\pm0.0534$ & $0.6406\pm0.0550$ & $0.6811\pm0.0463$                       \\
    ResNet+GMP & $0.6879\pm0.0380$ & $0.8368\pm0.0181$ & $0.6424\pm0.0729$ & $0.5567\pm0.1315$ & $0.5069\pm0.0739$ & $0.6357\pm0.0609$ & $0.6780\pm0.0330$                       \\
    ResNet+GMP+FC & $0.6991\pm0.0220$ & $0.8458\pm0.0137$ & $0.6748\pm0.0811$ & $0.5521\pm0.1230$ & $0.4925\pm0.4925$ & $0.6413\pm0.0440$ & $0.6890\pm0.0534$                    \\
    ResNet+GAP & $0.6965\pm0.0269$ & $0.8487\pm0.0131$ & $0.6777\pm0.0834$ & $0.5455\pm0.1240$ & $0.5019\pm0.0405$ & $0.6434\pm0.0552$ & $0.6927\pm0.6927$                      \\
    ResNet+GAP+FC & $0.7024\pm0.0287$ & $0.8471\pm0.0075$ & $0.6826\pm0.0890$ & $0.5556\pm0.1268$ & $0.5184\pm0.0523$ & $0.6509\pm0.0557$ & $0.6982\pm0.0427$                       \\
    
    \hline
    
    $FSConv$+FC & $0.7330\pm0.0303$ & $0.8395\pm0.0437$ & $0.6503\pm0.0229$ & $0.6964\pm0.0606$ & $0.5441\pm0.1294$ & $0.6826\pm0.0207$ & $0.6809\pm0.0273$                                     \\
    $FSConv$+GMP & $\mathbf{0.7622\pm0.0075}$ & $0.8766\pm0.0167$ & $\mathbf{0.7277\pm0.0228}$ & $\mathbf{0.7093\pm0.0540}$ & $0.5364\pm0.1062$ & $\mathbf{0.7125\pm0.0251}$ & $0.7328\pm0.0465$ \\
    $FSConv$+GMP+FC & $0.7286\pm0.0610$ & $0.8724\pm0.0341$ & $0.6955\pm0.0374$ & $0.6317\pm0.2017$ & $0.4529\pm0.0379$ & $0.6631\pm0.0592$ & $0.7200\pm0.0405$                                  \\
    $FSConv$+GAP & $0.5317\pm0.0886$ & $0.6830\pm0.0805$ & $0.3228\pm0.2408$ & $0.4418\pm0.2587$ & $0.3391\pm0.1835$ & $0.4467\pm0.1501$ & $0.4153\pm0.2376$                                    \\
    
    \hline
    \end{tabular}
    }
\end{table}

\begin{table}[htb]
    \centering
    \caption{Number of parameters and memory usage of the different CNN architectures tested for the patch-level Gleason grading task. KB: kilobytes.}
    \label{models_parameters}
    \begin{tabular}{|l|r|r|}
    \hline
    \multicolumn{1}{|c|}{\textbf{Experiment}} & \multicolumn{1}{|c|}{\textbf{Storage Space (KB)}} & \multicolumn{1}{|c|}{\textbf{Trainable Parameters}}  \\

    \hline
    \hline
    
    VGG19+FC        &  180700 & 46203652                                  \\
    VGG19+GMP        &   78290 & 19987716                                  \\
    VGG19+GMP+FC     &   79832 & 20380676                                  \\
    VGG19+GAP       &   78289 & 19987716                                  \\
    VGG19+GAP+FC    &   79833 & 20380676                                  \\
    
    \hline
    
    ResNet+FC       & 496022 & 126822916                                  \\
    ResNet+GMP       & 92579  &  23542788                                  \\
    ResNet+GMP+FC    & 97170  &  24716036                                  \\
    ResNet+GAP      & 92580  &  23542788                                  \\
    ResNet+GAP+FC   & 97179  &  24716036                                  \\
    
    \hline
    
    $FSConv$+FC       & 104899 & 26846212                                   \\
    $FSConv$+GMP       & 2486   &   630276                                   \\
    $FSConv$+GMP+FC    & 4026   &  1023236                                   \\
    $FSConv$+GAP      & 2485   &   630276                                   \\
    
    \hline
    \end{tabular}
\end{table}

Regarding the results obtained in the fine-tuned models, the use of architectures with residual blocks provided slightly worse results than the sequential approach, similarly as the previous results reported in the literature where sequential models used to outperform residual ones \cite{Esteban2019AProcesses, Arvaniti2018AutomatedLearning, Nir2018AutomaticExperts}. In relation to the use of different top models, no differences were found in the accuracy of the fine-tuned architectures, observing similar results for all of them.




In relation to $FSConv$ architecture, interesting results were obtained while testing the use of different top models. The best performing architecture to validate the system is the one with global-max pooling, $FSConv$+GMP. 
The outperforming of the global-max pooling compared to the fully-connected configuration could be explained by the reduction in the number of weights to be optimised (see Table \ref{models_parameters}), making the model simpler and more capable of generalising to new images, and by the invariance to the pattern location provided by the global-pooling operations. However, the $FSConv$ model did no converge properly using global-average poling in the top model ($FSConv$+GAP), an effect non observed in the case of fine-tuned architectures. The explanation of this behaviour could be related to the receptive field of the model. The receptive field is defined as the region of the image involved in the cross-correlation operation resulting in one output element in the activation map. As $FSConv$ is a shallow architecture, the final receptive field (i.e. in the last convolution layer) is limited, and then the extracted features are more local than the obtained by deeper architectures. Then, if the pattern to be detected is just located in a small portion of the tissue, the activation could be masked in the global average. This effect is not present in deep networks with a large receptive field as the VGG19 or ResNet, and it could explain the similar behaviour of both top models for the pre-trained networks. Therefore, the use of top models based on global-max pooling in shallow architectures allows to extract relevant features to train models from scratch reducing the number of trainable parameters of the model and increasing its robustness against size and location variability of the region of interest.

Paying attention to Table \ref{res1} and taking into account all the figures of merit, we conclude that $FSConv$+GMP configuration is the best performing one for the patch-level Gleason grading. In the validation set used, this model outperforms the VGG19+GMP+FC architecture in terms of accuracy ($0.7622$ compared to $0.7273$) and average F1-score ($0.7125$ against $0.6772$). Furthermore, the $FSConv$+GMP model performs specially well when distinguishing between GG3 and GG4, the most difficult task in the pathologists' work, reaching F1-scores of $0.7277$ and $0.7093$ respectively (see Table \ref{res1}). This is the first time in the literature that self-defined architectures trained from scratch outperform fine-tuned architectures from the state-of-the-art pre-trained in Imagenet for Gleason grading. Moreover, the reduced amount of parameters ($2\times10^{7}$ in the VGG19+GMP+FC model against $6\times10^{5}$ in the $FSConv$+GMP model, see Table \ref{models_parameters}), makes more convenient the $FSConv$ architecture for deployment. Thus, the model $FSConv$+GMP was trained using all the images in the cross-validation sets in order to evaluate its performance in the external test cohort.  

\begin{figure}[htb]
    \centering
      \subfloat[\label{cma}]{\includegraphics[width=.47\linewidth]{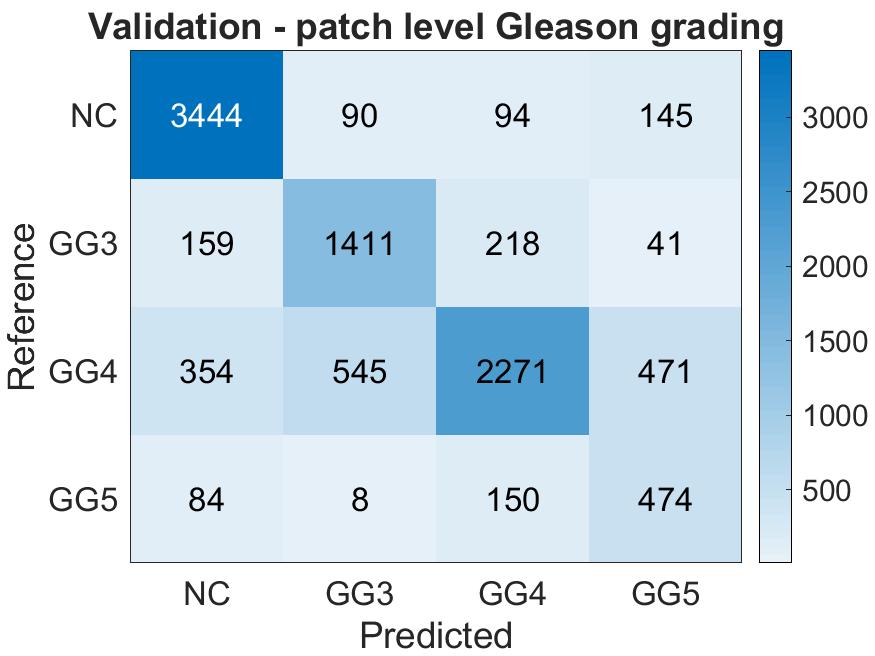}}
      \hspace*{\fill}
      \subfloat[\label{cmb}]{\includegraphics[width=.47\linewidth]{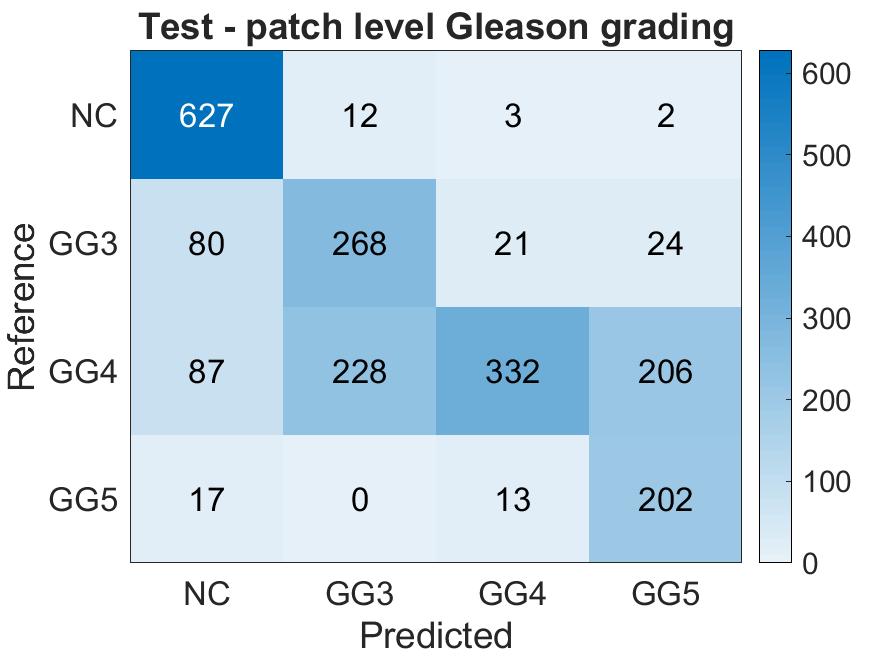}}
      \hspace*{\fill}
    \caption{Confusion Matrix of the patch-level Gleason grades prediction done by $FSConv$ network in (a) validation set and (b) test set.}
    \label{fig:cm}
\end{figure}

The results of the proposed model for the test set and a comparison of them with previous state-of-the-art works are reported in Table \ref{res2}. $\kappa$ value increases up to $0.77$ in the test subset for $FSConv$+GMP. In comparison with previous studies, our results outperform the state of the art, obtaining almost a strong agreement between our model and the pathologist, while just moderate agreement ($\kappa = 0.55$ \cite{Arvaniti2018AutomatedLearning}) was obtained previously in the test set. Figure \ref{fig:cm} shows the performance evaluation of $FSConv$. In particular, the confusion matrix for validation and test subsets are presented. From this figure, it can be observed that most of the errors occur between adjacent classes.

\begin{table}[htb]
\centering
\caption{Results for the patch-level Gleason grading in the test set for the model $FSConv$+GMP and comparison with previous literature. The metrics presented are accuracy (ACC), F1-Score (1S), computed per class and its average, and Cohen's quadratic kappa ($\kappa$). Note that for the results reported in previous literature not all the metrics were reported. GMP: global-max pooling.}
\label{res2}
\resizebox{\linewidth}{!}{
\begin{tabular}{|lc|c|cccc|c|c|}
\hline
\multicolumn{2}{|c|}{\textbf{Experiment}}                          & \textbf{ACC} & \multicolumn{4}{c|}{\textbf{F1S}} & \textbf{Avg-F1S} & \textbf{$\kappa$} \\
& &                                                              & NC & GG3 & GG4 & GG5 & &                                                        \\
\hline
\hline
$FSConv$+GMP & Test                                                 & $0.67$ & $0.86$ & $0.59$ & $0.54$ & $0.61$ & $0.65$ & $\textbf{0.77}$           \\
\hline
\multirow{ 2}{*}{Arvaniti et al.  \cite{Arvaniti2018AutomatedLearning}} & Validation  & - & - & - & - & - & - & $0.67$                                    \\
& Test  & -                                                      & - & - & - & - & - & $0.55$                                                      \\
\hline
Nir et al. \cite{Nir2019ComparisonImages} & Validation                  & - & - & - & - & - & - & $0.61$                                                  \\
\hline
\end{tabular}
}
\end{table}

\subsubsection{Model Interpretation}

One of the main drawbacks of deep learning models in medical practice is the lack of interpretability. This fact creates distrust in the clinicians, the final users of CAD systems. To deal with this problem, in this research we study the interpretability of the trained models by means of the Class Activation Maps technique (CAMs). Both VGG19+GMP+FC (the best fine-tuned model) and $FSConv$+GMP models are compared in this section using CAMs.



\begin{figure}[htb]
    \centering
    
    \subfloat{\resizebox{0.24\textwidth}{!}{
    \begin{tikzpicture}
    \node{\pgfimage[interpolate=true,height=5cm]{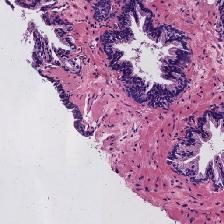}};
    \end{tikzpicture}}}
    \subfloat{\resizebox{0.24\textwidth}{!}{
    \begin{tikzpicture}
    \node{\pgfimage[interpolate=true,height=5cm]{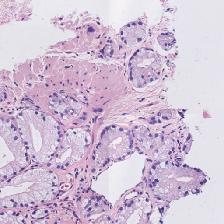}};
    \end{tikzpicture}}}
    \subfloat{\resizebox{0.24\textwidth}{!}{
    \begin{tikzpicture}
    \node {\pgfimage[interpolate=true,height=5cm]{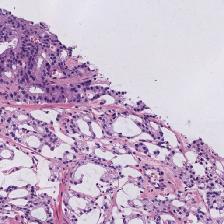}};
    \end{tikzpicture}}}
    \subfloat{\resizebox{0.24\textwidth}{!}{
    \begin{tikzpicture}
    \node {\pgfimage[interpolate=true,height=5cm]{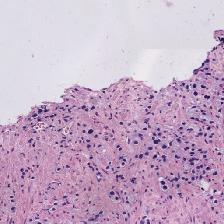}};
    \end{tikzpicture}}}

    \subfloat{\resizebox{0.24\textwidth}{!}{
    \begin{tikzpicture}
    \node{\pgfimage[interpolate=true,height=5cm]{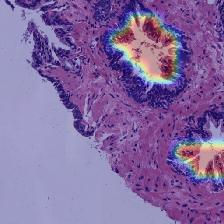}};
    \end{tikzpicture}}}
    \subfloat{\resizebox{0.24\textwidth}{!}{
    \begin{tikzpicture}
    \node{\pgfimage[interpolate=true,height=5cm]{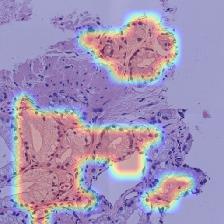}};
    \end{tikzpicture}}}
    \subfloat{\resizebox{0.24\textwidth}{!}{
    \begin{tikzpicture}
    \node {\pgfimage[interpolate=true,height=5cm]{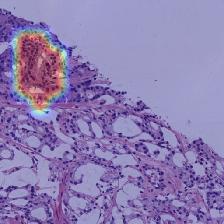}};
    \end{tikzpicture}}}
    \subfloat{\resizebox{0.24\textwidth}{!}{
    \begin{tikzpicture}
    \node {\pgfimage[interpolate=true,height=5cm]{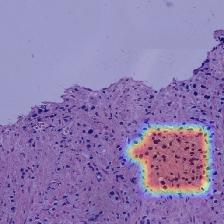}};
    \end{tikzpicture}}}

    \renewcommand{\thesubfigure}{a}
    \subfloat[\label{CAM1a}]{\resizebox{0.24\textwidth}{!}{
    \begin{tikzpicture}[spy using outlines={circle,yellow,magnification=4,size=3cm, connect spies}]
    \node {\pgfimage[interpolate=true,height=5cm]{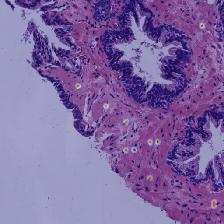}};
    \spy on (0.35,-0.7) in node [left] at (-1,-1);
    \end{tikzpicture}}}
    \renewcommand{\thesubfigure}{b}
    \subfloat[\label{CAM1b}]{\resizebox{0.24\textwidth}{!}{
    \begin{tikzpicture}[spy using outlines={circle,yellow,magnification=1.5,size=2.5cm, connect spies}]
    \node {\pgfimage[interpolate=true,height=5cm]{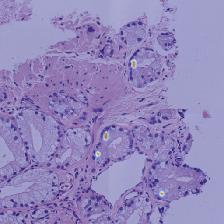}};
    \spy on (0.7,1) in node [left] at (-1,1);
    \end{tikzpicture}}}
    \renewcommand{\thesubfigure}{c}
    \subfloat[\label{CAM1c}]{\resizebox{0.24\textwidth}{!}{
    \begin{tikzpicture}[spy using outlines={circle,yellow,magnification=2.2,size=2.5cm, connect spies}]
    \node {\pgfimage[interpolate=true,height=5cm]{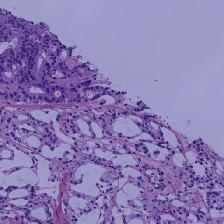}};
    \spy on (-1.2,0.5) in node [right] at (1,-1);
    \end{tikzpicture}}}
    \renewcommand{\thesubfigure}{d}
    \subfloat[\label{CAM1d}]{\resizebox{0.24\textwidth}{!}{
    \begin{tikzpicture}[spy using outlines={circle,yellow,magnification=2,size=2cm, connect spies}]
    \node {\pgfimage[interpolate=true,height=5cm]{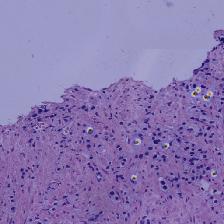}};
    \spy on (2,0.5) in node [left] at (-1,1);
    \end{tikzpicture}}}
    \caption{Original image (first row) and Class Activation Maps (CAMs) obtained by the VGG19 model (second row) and the $FSConv$ network (third row) in four images correctly classified. Non-Cancerous (a), Gleason grade $3$ (b),  Gleason grade $4$ (c) and Gleason grade $5$ (d).}
    \label{fig:CAM1}
\end{figure}

\begin{figure}[htb]
    \centering
    
    \subfloat{\resizebox{0.24\textwidth}{!}{
    \begin{tikzpicture}
    \node{\pgfimage[interpolate=true,height=5cm]{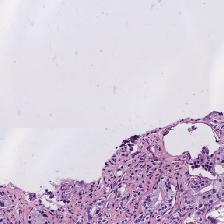}};
    \end{tikzpicture}}}
    \subfloat{\resizebox{0.24\textwidth}{!}{
    \begin{tikzpicture}
    \node{\pgfimage[interpolate=true,height=5cm]{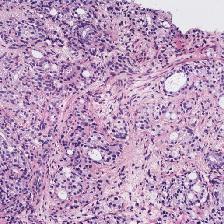}};
    \end{tikzpicture}}}
    \subfloat{\resizebox{0.24\textwidth}{!}{
    \begin{tikzpicture}
    \node {\pgfimage[interpolate=true,height=5cm]{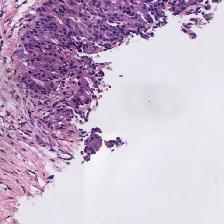}};
    \end{tikzpicture}}}

    \subfloat{\resizebox{0.24\textwidth}{!}{
    \begin{tikzpicture}
    \node{\pgfimage[interpolate=true,height=5cm]{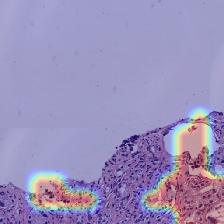}};
    \end{tikzpicture}}}
    \subfloat{\resizebox{0.24\textwidth}{!}{
    \begin{tikzpicture}
    \node{\pgfimage[interpolate=true,height=5cm]{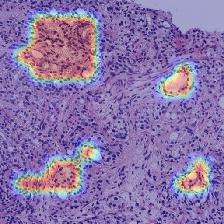}};
    \end{tikzpicture}}}
    \subfloat{\resizebox{0.24\textwidth}{!}{
    \begin{tikzpicture}
    \node {\pgfimage[interpolate=true,height=5cm]{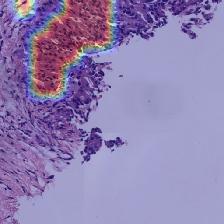}};
    \end{tikzpicture}}}

    \renewcommand{\thesubfigure}{a}
    \subfloat[\label{CAM2a}]{\resizebox{0.24\textwidth}{!}{
    \begin{tikzpicture}[spy using outlines={circle,yellow,magnification=2,size=2cm, connect spies}]
    \node {\pgfimage[interpolate=true,height=5cm]{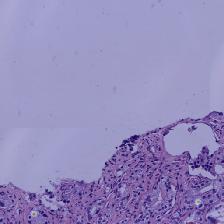}};
    \spy on (-1.8,-2.25) in node [left] at (0,1);
    \end{tikzpicture}}}
    \renewcommand{\thesubfigure}{b}
    \subfloat[\label{CAM2b}]{\resizebox{0.24\textwidth}{!}{
    \begin{tikzpicture}[spy using outlines={circle,yellow,magnification=3,size=2.5cm, connect spies}]
    \node {\pgfimage[interpolate=true,height=5cm]{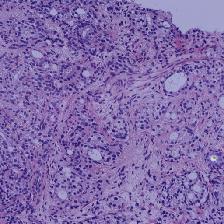}};
    \spy on (2.3,-1) in node [left] at (-0,0.75);
    \end{tikzpicture}}}
    \renewcommand{\thesubfigure}{c}
    \subfloat[\label{CAM2c}]{\resizebox{0.24\textwidth}{!}{
    \begin{tikzpicture}[spy using outlines={circle,yellow,magnification=2,size=2cm, connect spies}]
    \node {\pgfimage[interpolate=true,height=5cm]{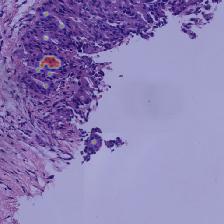}};
    \spy on (-1.4,1) in node [right] at (0.4,-1.4);
    \end{tikzpicture}}}
    
    \caption{Original images (first row) and Class Activation Maps (CAMs) obtained on the VGG19 model (second row) and the $FSConv$ network (third row) in images with GG4 correctly classified by the $FSConv$. The VGG19 model classification of those cases is GG3 in (a) and (b) and GG5 in (c).}
    \label{fig:CAM2}
\end{figure}

 This technique was proposed in \cite{Zhou2016LearningLocalization} as a procedure to obtain a heatmap indicating the regions of the input image to which the model is paying attention to predict certain class. CAMs for both models are obtained for images correctly classified (see Figure \ref{fig:CAM1}) and for images miss-classified by the VGG19 model (see Figure \ref{fig:CAM2}). These illustrations are organised as follows: the first row corresponds to the original patch, and the second and third rows show the CAMs for VGG19 and  $FSConv$ models, respectively. In Figure \ref{fig:CAM1} each column shows an example per class: NC, GG3, GG4 and GG5 accordingly. The main difference in the results obtained by VGG19 and $FSConv$ is the best differentiation between GG3 and GG4 by the second model (see Table \ref{res1}), the most difficult task in the pathologists' work. In Figure \ref{fig:CAM2} three of those cases are presented in each column: two cases predicted by the VGG19 as GG3 and one as GG5, respectively. Those cases were correctly classified as GG4 by $FSConv$ model.

CAMs obtained for VGG19 in NC, GG3 and GG4 show that the model is basing the decision in glandular regions detected and classified correctly. In the case of GG5, the highlighted region presents a group of single cells and infiltrating cords without lumen formation, characteristic patterns of poor differentiate tissue in GG5. In the case of $FSConv$ architecture, the CAM heatmap does not detect large regions, but small dots instead. Although the glandular regions are not detected, paying attention to the position where the dots are pointing at, we can extract interesting insights (see Figure \ref{fig:CAM1}). In the case of GG4, the map is activated in a small nest belonging to a fused-glands structure with irregular cribriform shape. Regarding the GG3 image, the dot indicates thick cytoplasm in different medium-sized tubular glands. In the image marked as GG5, the CAM highlights single isolated cells with hyperchromasia. Less interpretable is the CAM obtained in the NC image, where any gland is detected. We speculate that the model carries out this classification by dismissing the presence of cancerous patterns. Regarding the cases where VGG19 miss-classifies GG4 in Figure \ref{fig:CAM2}, a correct detection of the regions of interest is observed. However, these glandular regions are not correctly classified as GG4, while $FSConv$ model does it just paying attention to closed lumens in small ill-formed glands. At this stage of understanding, we believe that this fact is the cause of the different performance by both models. VGG19 focuses the prostate cancer detection on detecting epithelial and glandular regions, and these structures present a larger heterogeneity than its basic components (colour and size of individual glands, diameter and opening degree of lumens in the glandular region, etc.). This could be the reason why the VGG19 generalises slightly worse than $FSConv$.

\subsubsection{Validation on External Databases}

\begin{figure}[htb]
    \centering
    
      \subfloat[\label{figarva}]{\includegraphics[width=.22\linewidth]{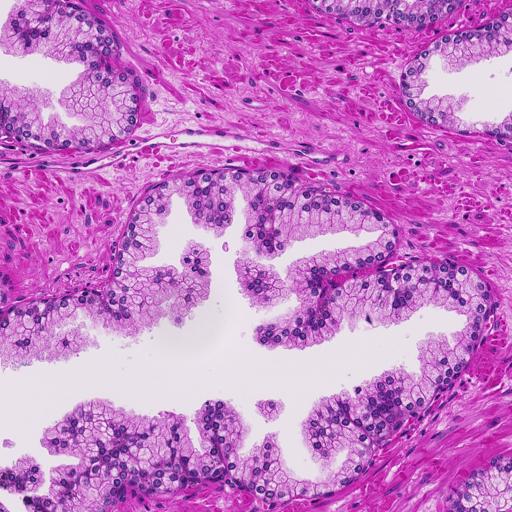}}
      \hspace*{\fill}
      \subfloat[\label{figarvb}]{\includegraphics[width=.22\linewidth]{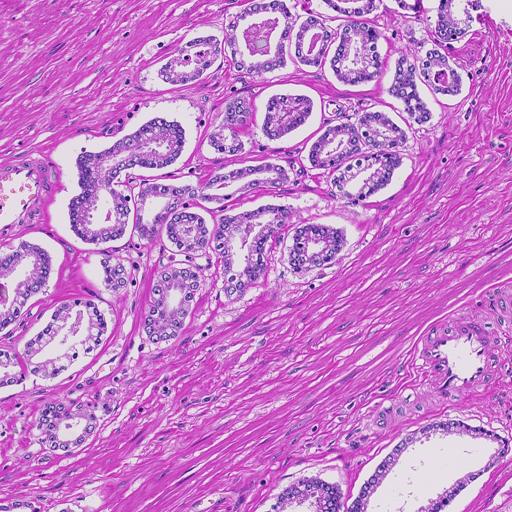}}
      \hspace*{\fill}
      \subfloat[\label{figarvc}]{\includegraphics[width=.22\linewidth]{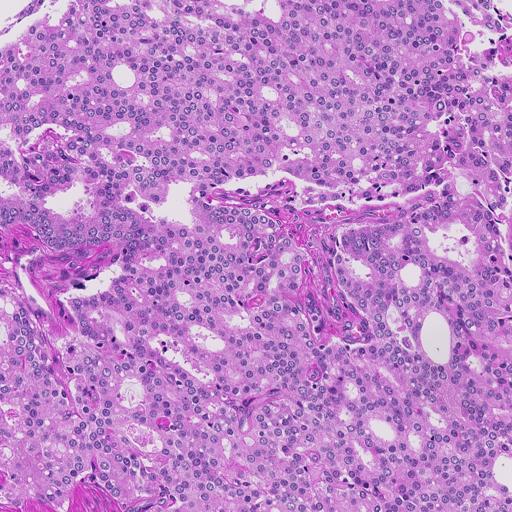}}
      \hspace*{\fill}
      \subfloat[\label{figarvd}]{\includegraphics[width=.22\linewidth]{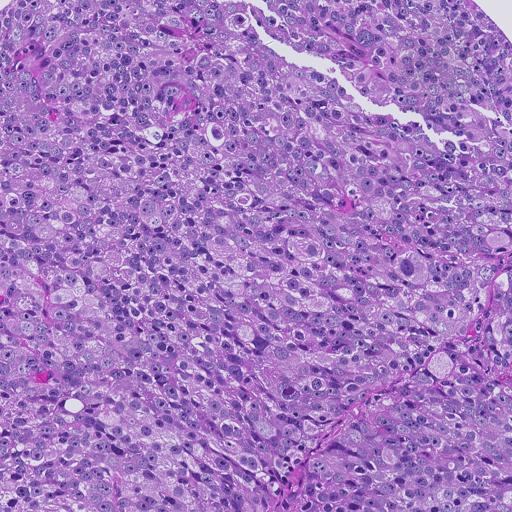}}
      \hspace*{\fill}
      
      \subfloat[\label{figgera}]{\includegraphics[width=.22\linewidth]{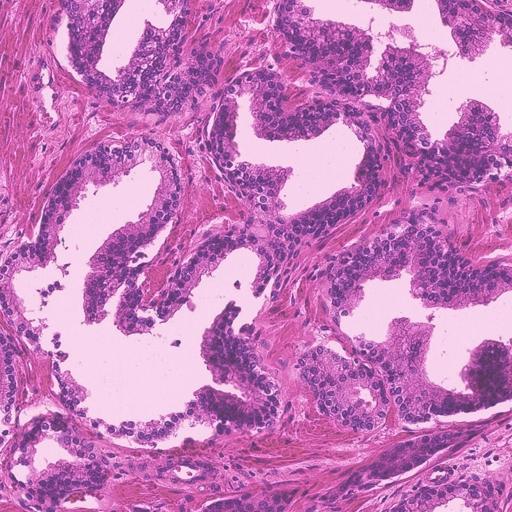}}
      \hspace*{\fill}
      \subfloat[\label{figgerb}]{\includegraphics[width=.22\linewidth]{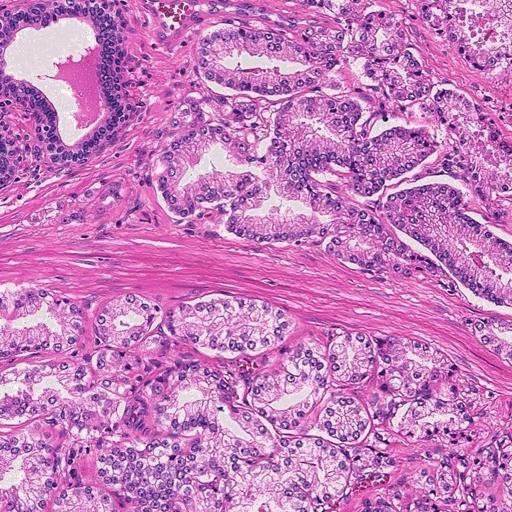}}
      \hspace*{\fill}
      \subfloat[\label{figerc}]{\includegraphics[width=.22\linewidth]{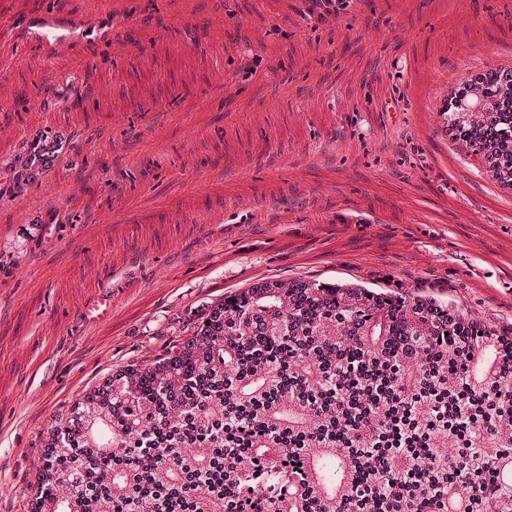}}
      \hspace*{\fill}
      \subfloat[\label{figgerd}]{\includegraphics[width=.22\linewidth]{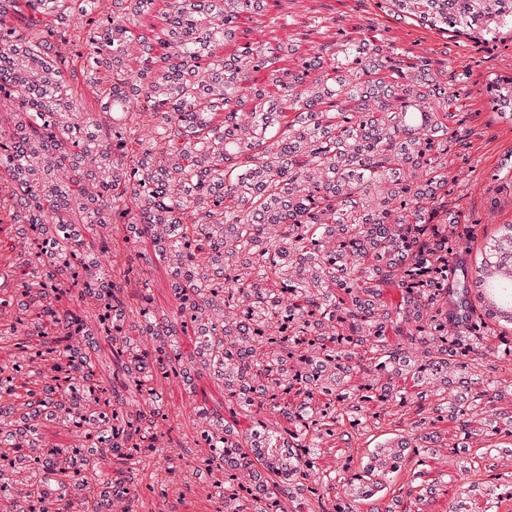}}
      \hspace*{\fill}
      
    \caption{Examples of patches used from the external database from Arvaniti et al. (first row) and Gerytch et al. (second row). (a) and (e): Benign glands; (b) and (f): Patches containing GG3 patterns; (c) and (d): Patches containing GG4 patterns; (d) and (h): Patches containing GG5 patterns.}
    \label{fig:images_arvaniti}
\end{figure}

\begin{figure}[htb]
    \centering
      \subfloat[\label{cmArvanitia}]{\includegraphics[width=.47\linewidth]{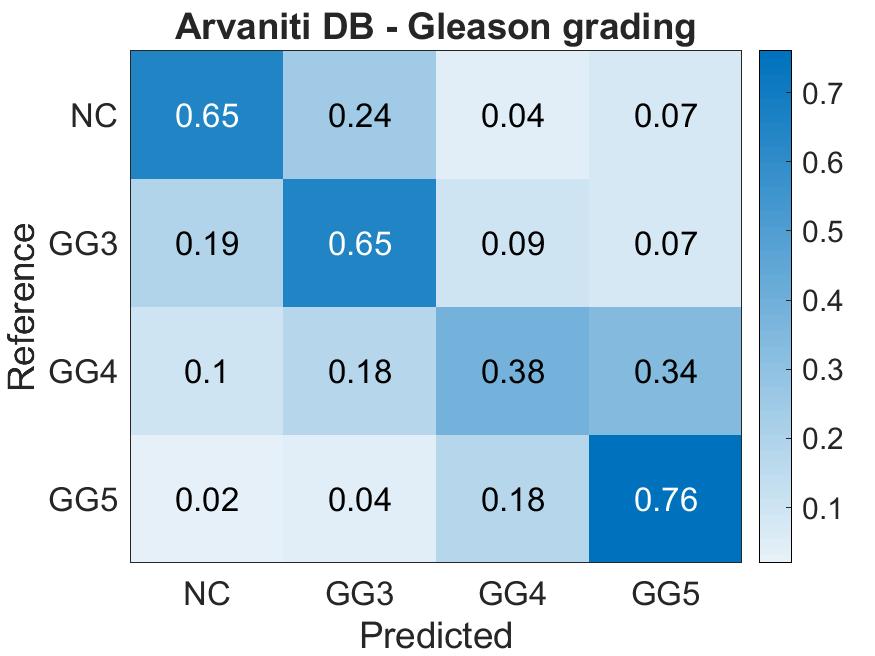}}
      \hspace*{\fill}
      \subfloat[\label{cmArvanitib}]{\includegraphics[width=.47\linewidth]{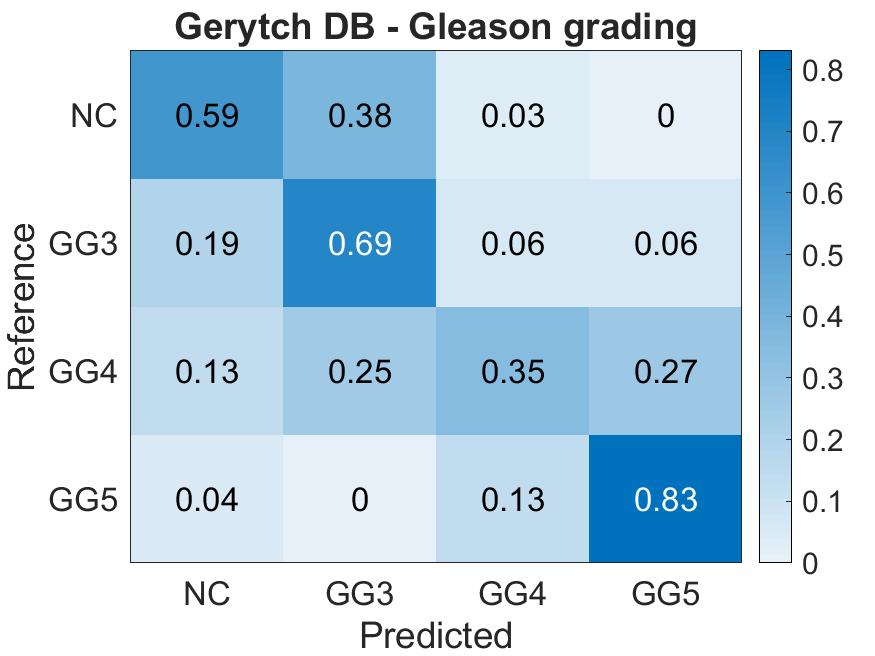}}
      \hspace*{\fill}
    \caption{Confusion Matrix of the patch-level Gleason grades prediction in external databases using the proposed $FSConv$ model. (a): Arvaniti database and (b): Gerytch database.}
    \label{fig:cmArvaniti}
\end{figure}

With the purpose of testing the generalization capability of the trianed model, $FSConv$ net was validated on two external databases. The databases used were shared by Arvaniti et al. \cite{Arvaniti2018AutomatedLearning} and Gerytch et al. \cite{Gertych2015MachineProstatectomies}. The first database is composed of $886$ cores from Tissue-Micro Arrays digitised at $40\times$ magnification, and the second has $625$ patches of prostate histology images at $20\times$ magnification. Each core was resized to $10\times$ resolution and a central patch with dimensions $512^{2}$ was extracted. For both databases, the ground truth was generated following the procedure in \cite{Arvaniti2018AutomatedLearning}. Non-cancerous patches were extracted from images with only benign structures annotated, labels GG3, GG4, and GG5 were assigned to patches with only the corresponding grade annotated. Examples of the obtained images from the Arvaniti et al. and Gerytch et al. databases are presented in the first and second rows of Figure \ref{fig:images_arvaniti}, respectively. Note that the H\&E stain color images are different from those appearing in the SICAPv2 database (see Figure \ref{fig1} for examples of the images used to train the developed models). To normalise the colour distribution of the images in external databases, the method presented in \cite{Vahadane2015Structure-preservedImages} was used after applying a channel-wise histogram matching of the external images to a SICAPv2 database reference image. This image was selected by the expert pathologists involved in this work based on its structural and colour properties. Then, our best performing model, i.e. $FSConv$, was used to predict and evaluate our performance on the external databases. Table \ref{res_arvaniti} and Figure \ref{fig:cmArvaniti} show the obtained figures of merit and confusion matrices, respectively.

\begin{table}[htb]
    \centering
    \caption{Results of the patch-level Gleason grading in the Arvaniti and Gerytch databases by our proposed model, $FSConv$. The metrics presented are accuracy (ACC), F1-Score (F1S), computed per class and its average, and Cohen's quadratic kappa ($\kappa$).}
    \label{res_arvaniti}
    \resizebox{\linewidth}{!}{
    \begin{tabular}{|l|c|cccc|c|c|}
    \hline
    \multicolumn{1}{|c|}{\textbf{Database}} & \textbf{ACC} & \multicolumn{4}{|c|}{\textbf{F1S}} & \textbf{Avg-F1S} & \textbf{$\kappa$}\\
     & & NC & GG3 & GG4 & GG5 & & \\
    \hline
    \hline
    
    Arvaniti et al. \cite{Arvaniti2018AutomatedLearning} & $0.5861$ & $0.5660$ & $0.6858$ & $0.4688$ & $0.5603$ & $0.5702$ & $0.6410$\\
    Gerytch et al. \cite{Gertych2015MachineProstatectomies} & $0.5136$ & $0.2901$ & $0.6162$ & $0.4990$ & $0.4958$ & $0.4753$ & $0.5116$\\

    \hline
    \end{tabular}
    }
\end{table}

The obtained results in Arvaniti et al. database were slightly worse than the ones reached in our test cohort. The macro-averaged F1 score was $0.57$, while $0.65$ was obtained in the test cohort (see Table \ref{res2}). To the best of the authors's knowledge, this is the first time in the literature that a model trained for patch-level Gleason grading in tested on an external database. This is a challenging task, taking into account the known inter-pathologist variability of the Gleason grading task and the differences in the histology sample preparation. Thus, the difference in the results could be explained by those factors. In comparison to the results obtained in \cite{Arvaniti2018AutomatedLearning} on this database, the reported $\kappa$ in the test subset was $0.55$ (see Table \ref{res2}), while the $\kappa$ obtained by our model was $0.64$. Our proposed model outperforms the current state of the art on this set of images, even though we used the whole database for testing, and they reported the result on a specific test subset.

Regarding the obtained results on the Gerytch et al. database, a macro-averaged F1 score of $0.47$, and a $\kappa$ of $0.51$ were obtained. Note that the small amount of non cancerous patches in this database ($32$ patches with only benign annotation, compared to $116$ in Arvaniti et al. set) could be negatively affecting the figures of merit. Unfortunately, to the best of the authors' knowledge, no work has been reported on the use of the entire set of grades on this database, which makes the comparison impossible.

\subsection{Cribriform Pattern Detection}
\label{cribriform}

\begin{figure}[htb]
    \centering
      \subfloat[\label{crib1a}]{\includegraphics[width=.45\linewidth]{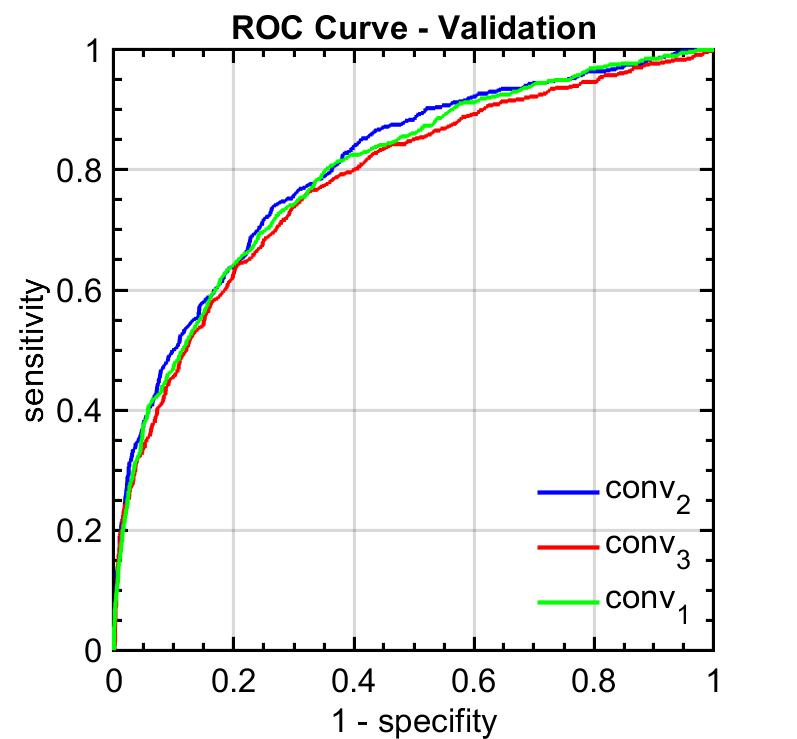}}
      \hspace*{\fill}
      \subfloat[\label{crib1b}]{\includegraphics[width=.45\linewidth]{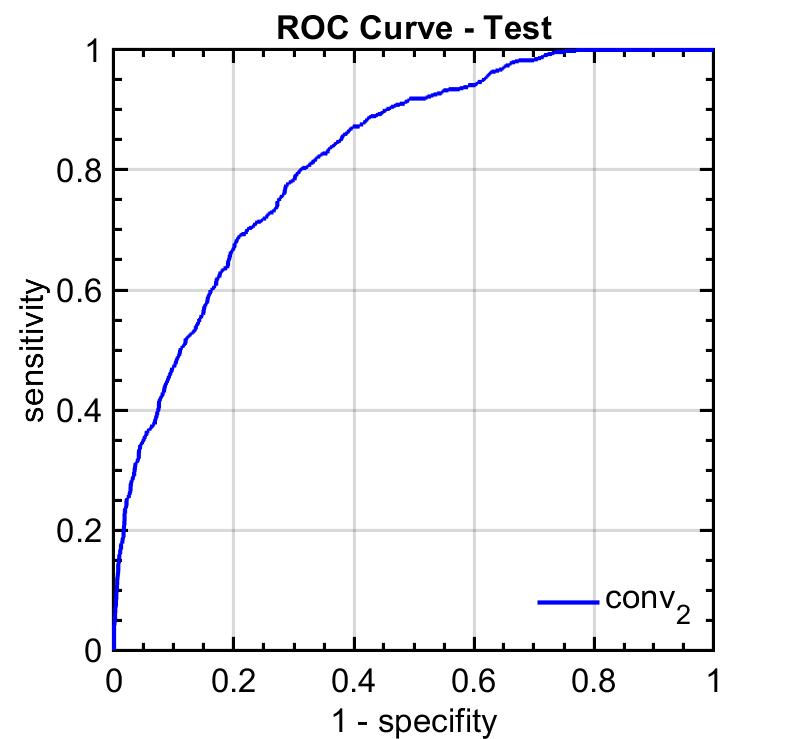}}
      \hspace*{\fill}
    \caption{ROC curves obtained for cribriform pattern detection in samples with Gleason grade $4$.}
    \label{fig:crib1}
\end{figure}

To detect cribriform patterns in GG4 patches, $FSConv$ trained in the Gleason grading stage was re-trained as specified in the \autoref{chap:crib} with a learning rate of $0.001$ and a batch size of $32$ samples during $200$ epochs. The results were optimised freezing the weights of the convolutional filters at different depths. Concretely, at filters $conv_{1}$, $conv_{2}$ and $conv_{3}$ (see Table \ref{gleasonNet1} for $FSConv$ architecture details). The output probability of each model was used to compute the Receiver Operative Curve (ROC) and evaluate the Area Under Curve (AUC). Then, probabilities were thresholded to output a positive classification when they are above $50\%$. The results obtained for the cross-validation set are presented in Table \ref{resCrib1}, and the Receiver-Operative-Curve in Figure \ref{fig:crib1} (a).

\begin{table}[htb]
    \centering
    \caption{Results in the detection of cribriform pattern in the validation set. The accuracy (ACC), Sensitivity, specificity and area under ROC curve (AUC) are presented for the fine-tuned $FSConv$ model freezing up to the convolutional layers $conv_{1}$, $conv_{2}$ or $conv_{3}$.}
    \label{resCrib1}
    \resizebox{\linewidth}{!}{
    \begin{tabular}{|l|c|c|c|c|}
    \hline
    \textbf{Experiment} & \textbf{ACC} & \textbf{Sensitivity} & \textbf{Specificity} & \textbf{AUC}\\
    \hline
    \hline
    
    $conv_{1}$ & $0.8218\pm0.0541$          & $0.8837\pm0.0525$             & $0.5263\pm0.1159$           & $0.8172\pm0.0689$            \\
    $conv_{2}$ & $\mathbf{0.8350\pm0.0599}$ & $\mathbf{0.8993\pm0.0436}$    & $0.5223\pm0.1435$           & $\mathbf{0.8225\pm0.0733}$   \\
    $conv_{3}$ & $0.8103\pm0.0712$          & $0.8586\pm0.0650$             & $\mathbf{0.5476\pm0.2229}$  & $0.7965\pm0.1018$            \\

    \hline
    \end{tabular}
    }
\end{table}

\begin{figure}[htb]
    \centering
      \subfloat[\label{crib2a}]{\includegraphics[width=.32\linewidth]{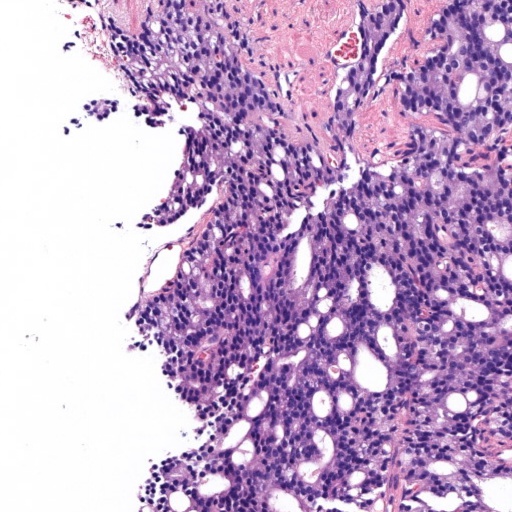}}
      \hspace*{\fill}
      \subfloat[\label{crib2b}]{\includegraphics[width=.32\linewidth]{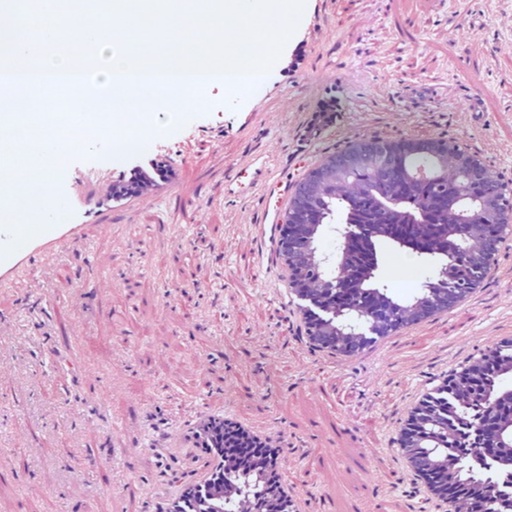}}
      \hspace*{\fill}
      \subfloat[\label{crib2c}]{\includegraphics[width=.32\linewidth]{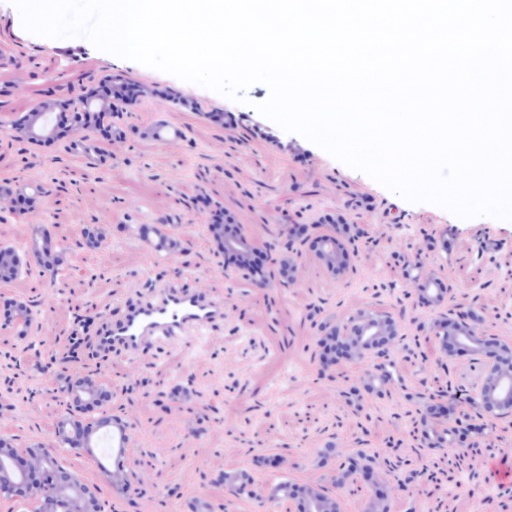}}
      \hspace*{\fill}
      
      \subfloat[\label{crib2d}]{\includegraphics[width=.32\linewidth]{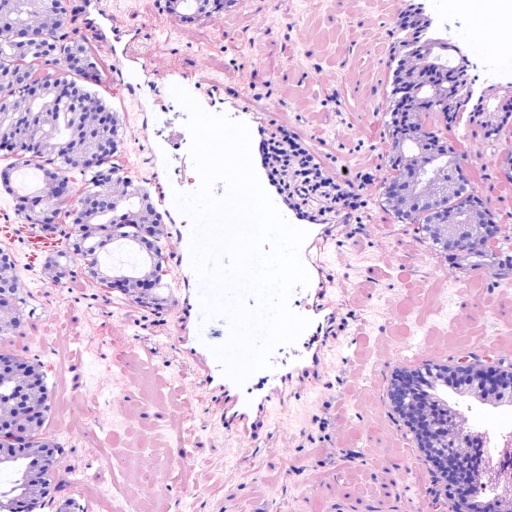}}
      \hspace*{\fill}
      \subfloat[\label{crib2e}]{\includegraphics[width=.32\linewidth]{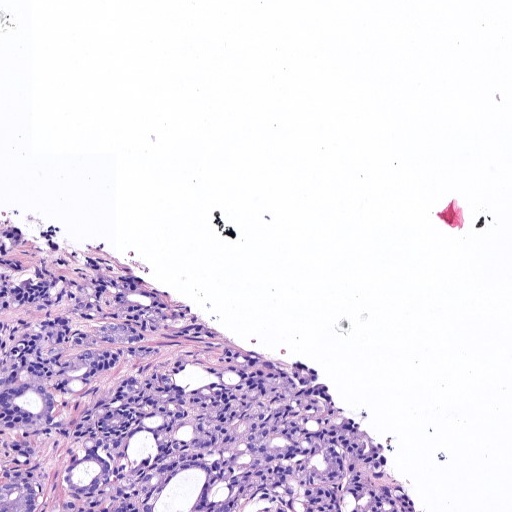}}
      \hspace*{\fill}
      \subfloat[\label{crib2f}]{\includegraphics[width=.32\linewidth]{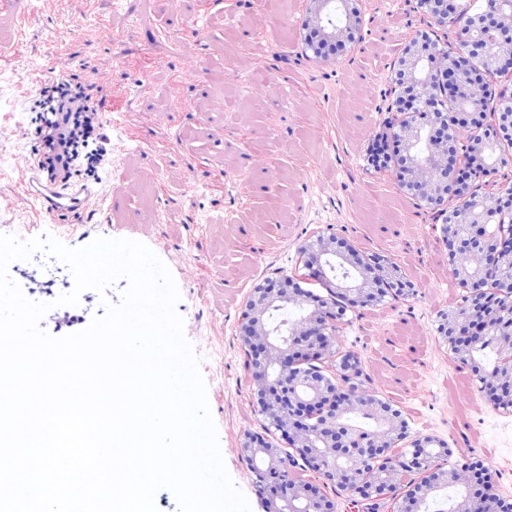}}
      \hspace*{\fill}
      
    \caption{Examples of the system performance in the test subset for cribriform pattern detection. (a): True Positive, (b): True Positive, (c): True Negative, (d): False Positive, (e): False Positive, (f): False Positive.}
    \label{fig:crib2}
\end{figure}

The best results were obtained for the validation set by the network whose weights were frozen up to the layer $conv_{2}$. Thus, just the last layer, $conv_{3}$ and the output neuron were trained. The accuracy obtained through this configuration was $0.8225$, with a sensitivity and specificity of $0.8993$ and $0.5223$, respectively. The reached AUC was $0.8225$. Slightly better results were obtained by this model in the test subset. The ROC computed in the test subset is presented in Figure \ref{fig:crib1} (b), and it encloses an AUC of $0.8240$. This value is at the permissible confidence level of systems for medical applications, above $0.80$ \cite{Swets1988MeasuringSystems}. Although the accuracy value decreases to $0.7239$, the sensitivity and specificity are more balanced, with values $0.7168$ and $0.7586$, respectively. To the best of the authors's knowledge, this is the first time that the detection of cribriform patterns in histology prostate images is addressed and evaluated, so that it is not possible to establish comparison with previous works. Nevertheless, the studies comparing the inter-observer variability of the Gleason patterns classification show the challenging character of this task. In \cite{Kweldam2016GleasonPathologists} the reproducibility in this problem was studied with $23$ genitourinary pathologists. The consensus was achieved for cribriform glands in only 23\% of the cases, and a consensus was not reached in how to classify the complex fused glands with cribriform shapes. We observed that the misclassified instances in our approach were mainly due to this kind of pattern. In Figure \ref{fig:crib2} few representative examples are presented, being (d), (e), and (f) images with complex fused glands that the model misclassified as  cribriform pattern. Therefore, the results obtained by the model are auspicious, and its main limitation is the misclassification of patterns with large inter-pathologist variability.

\subsection{WSI-Level Gleason Scoring}
\label{region}


Once the patch-level prediction is performed with model $FSConv$, the probability maps for each Gleason grade are obtained, as specified in the \autoref{chap:WSIscoring}. The usability of these maps in the clinical practice were qualitatively validated by expert pathologists with satisfactory results.

\begin{figure}[htb]
    \centering
    \renewcommand{\thesubfigure}{a}
    \subfloat[\label{region1a}]{\resizebox{0.40\textwidth}{!}{
    \begin{tikzpicture}[spy using outlines={circle,yellow,magnification=2.5,size=2.5cm, connect spies}]
    \node {\pgfimage[interpolate=true,height=5cm]{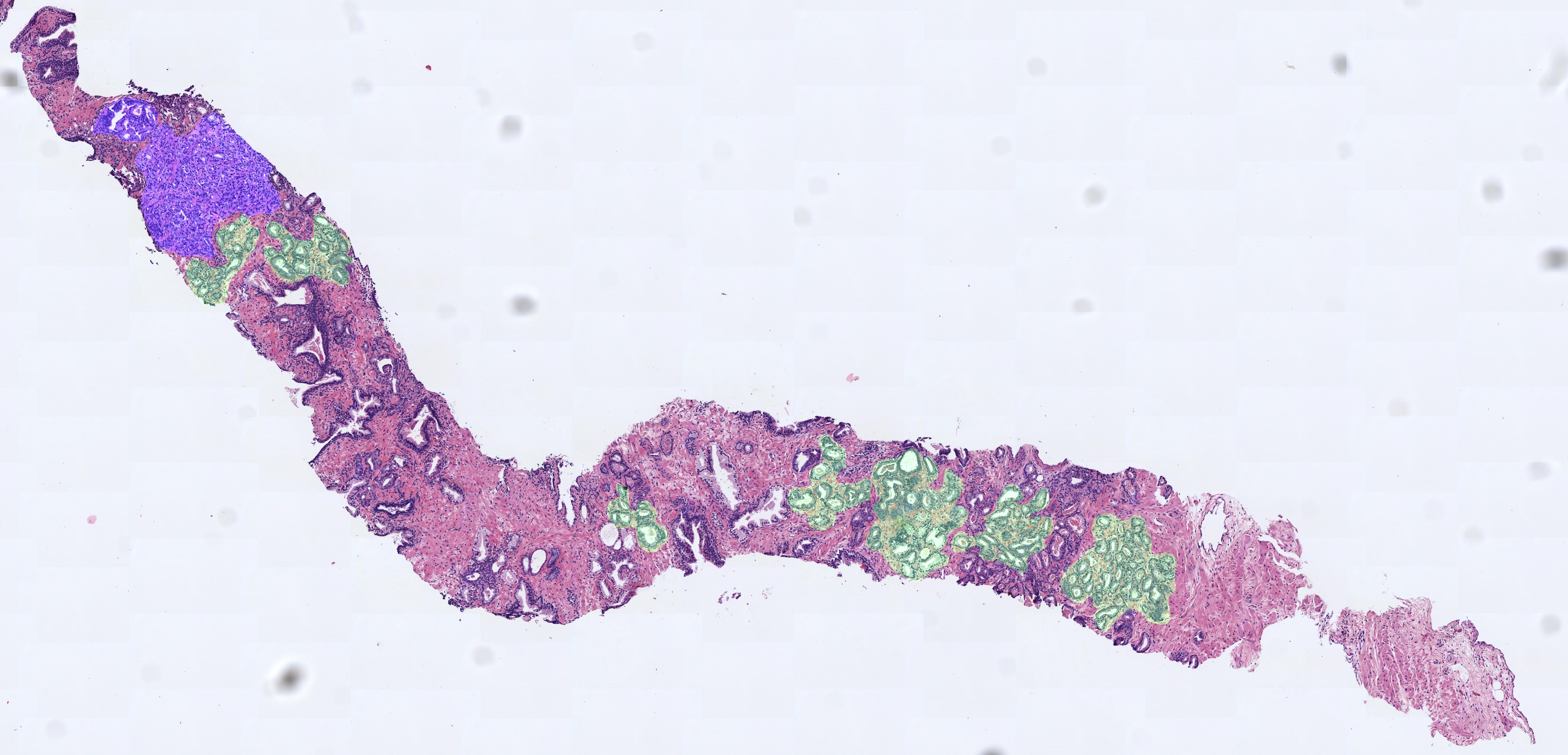}};
    \spy on (-1,-1) in node [left] at (4,1.5);
    \end{tikzpicture}}}
    \renewcommand{\thesubfigure}{c}
    \subfloat[\label{region1c}]{\resizebox{0.40\textwidth}{!}{
    \begin{tikzpicture}[spy using outlines={circle,yellow,magnification=10,size=3cm, connect spies}]
    \node {\pgfimage[interpolate=true,height=5cm]{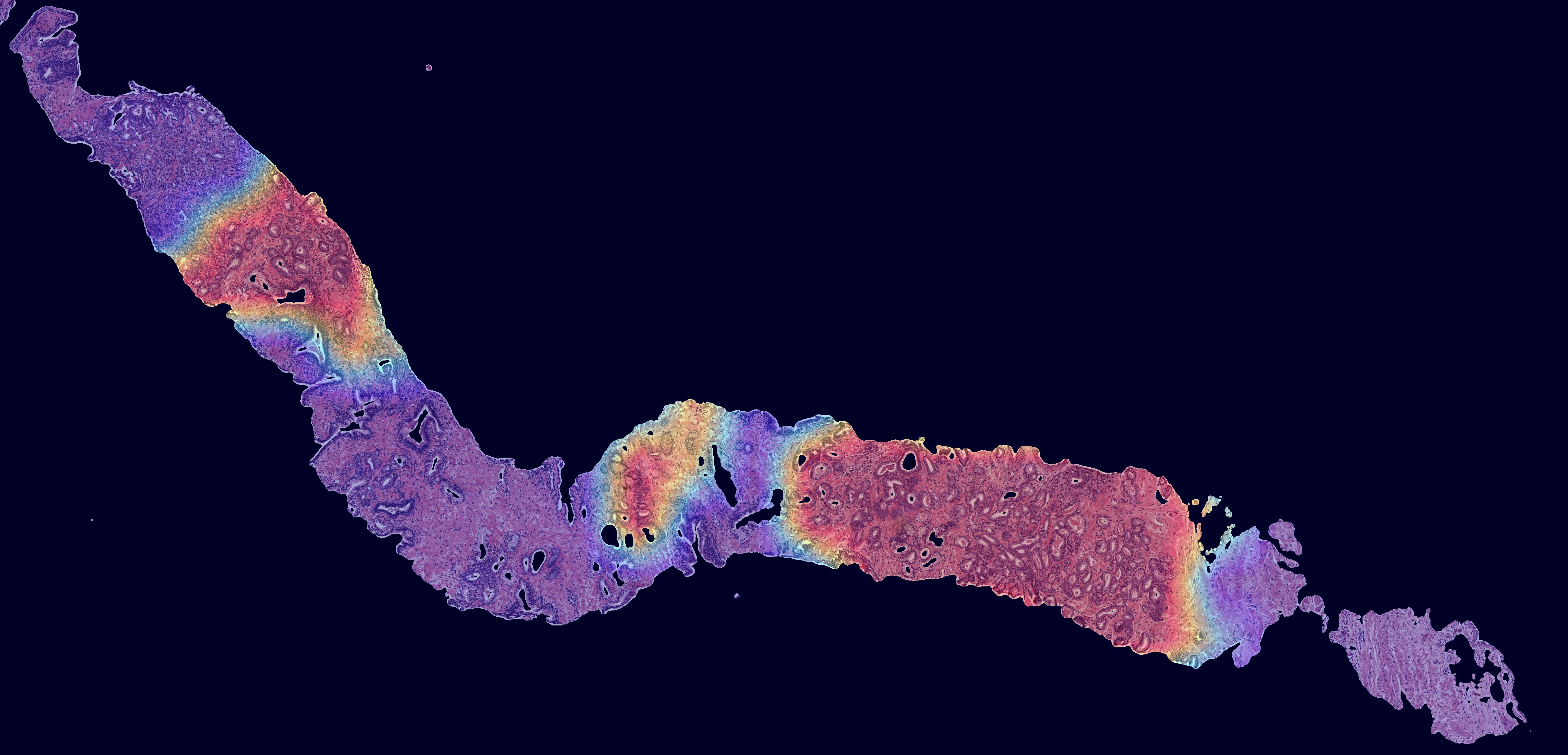}};
    \spy on (-3.1,0.9) in node [left] at (1,1.5);
    \spy on (1.4,-1.1) in node [left] at (4.5,1.5);
    \end{tikzpicture}}}
    
    \renewcommand{\thesubfigure}{b}
    \subfloat[\label{region1b}]{\resizebox{0.40\textwidth}{!}{
    \begin{tikzpicture}[spy using outlines={circle,yellow,magnification=2.5,size=2.5cm, connect spies}]
    \node {\pgfimage[interpolate=true,height=5cm]{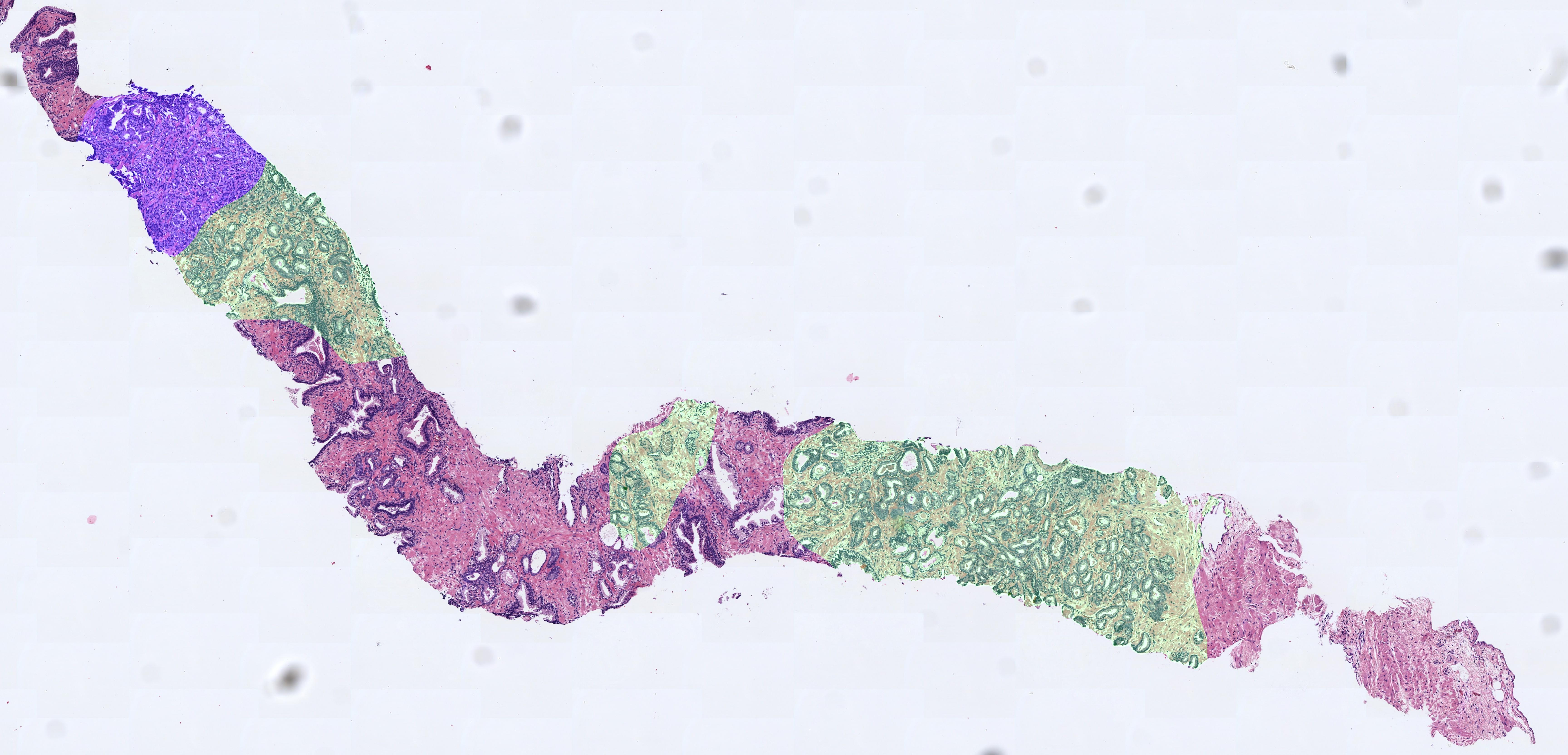}};
    \spy on (-2.5,-0.7) in node [left] at (1,1.5);
    \spy on (-1,-1) in node [left] at (4,1.5);
    \end{tikzpicture}}}
    \renewcommand{\thesubfigure}{d}
    \subfloat[\label{region1d}]{\resizebox{0.40\textwidth}{!}{
    \begin{tikzpicture}[spy using outlines={circle,yellow,magnification=10,size=3cm, connect spies}]
    \node {\pgfimage[interpolate=true,height=5cm]{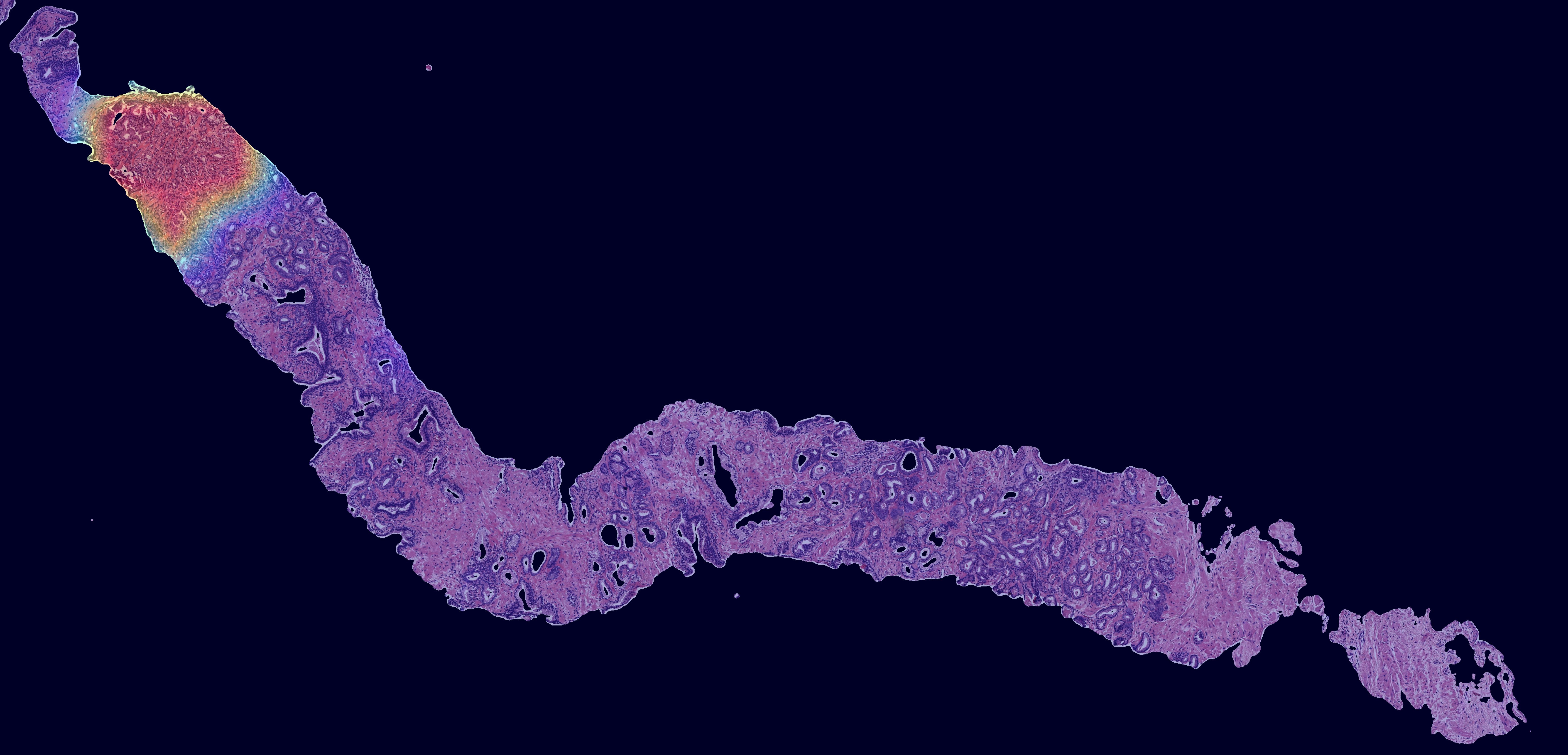}};
    \spy on (-4,1.5) in node [left] at (1,1.5);
    \end{tikzpicture}}}

    \subfloat{\resizebox{0.40\textwidth}{!}{
    }}
    \renewcommand{\thesubfigure}{e}
    \subfloat[\label{region1e}]{\resizebox{0.40\textwidth}{!}{
    \begin{tikzpicture}
    \node {\pgfimage[interpolate=true,height=5cm]{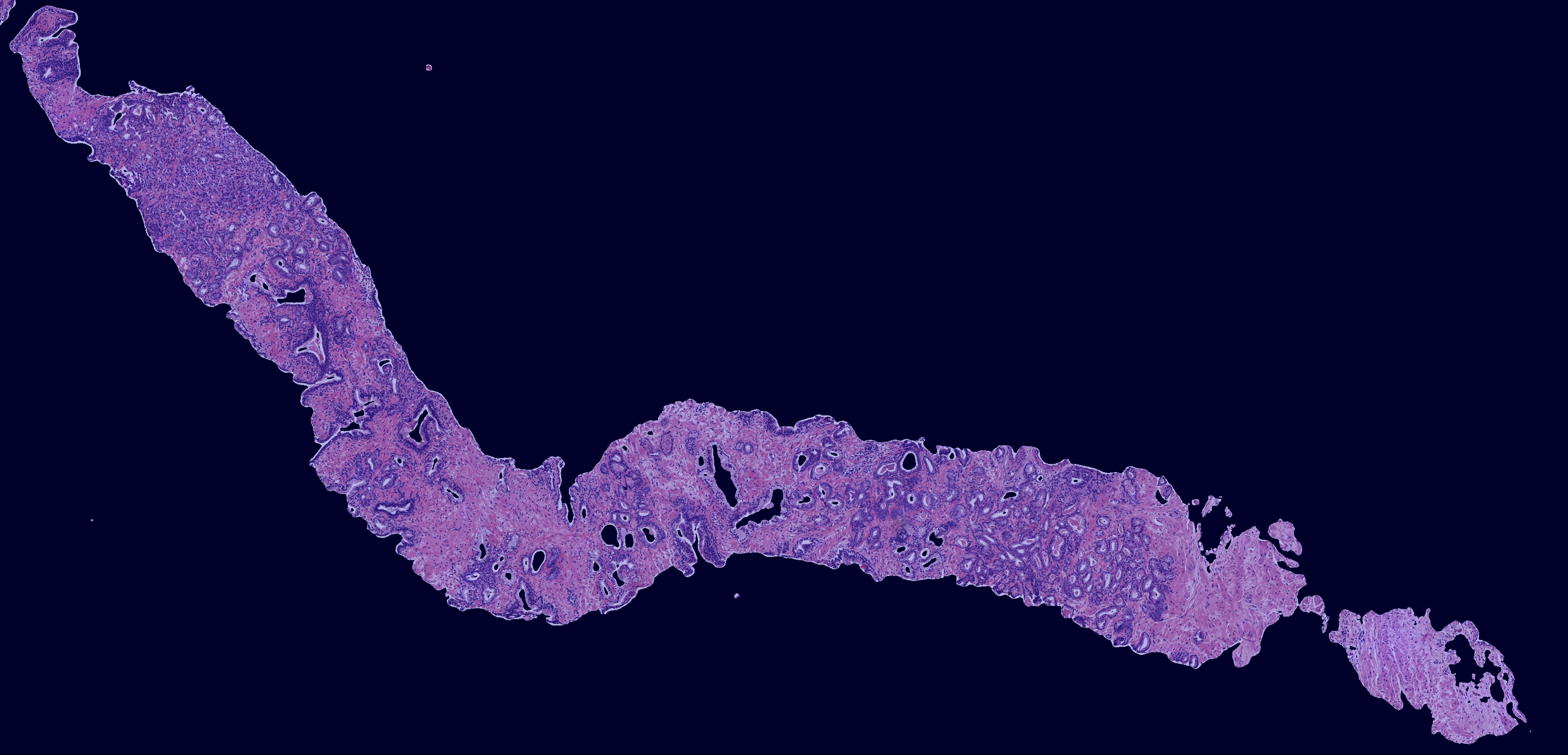}};
    \end{tikzpicture}}}

    \caption{Whole slide image level prediction of a biopsy diagnosed as Gleason Score $3+4=7$. (a): manual annotations, (b): system predictions. Green:  GG3, Blue:  GG4, red: GG5. (c): GG3 heatmap, (d): GG4 heatmap, (e): GG5 heatmap.}
    \label{fig:region1}
\end{figure}

\begin{figure}
    \centering
    
    \renewcommand{\thesubfigure}{a}
    \subfloat[\label{region2a}]{\resizebox{0.40\textwidth}{!}{
    \begin{tikzpicture}
    \node {\pgfimage[interpolate=true,height=5cm]{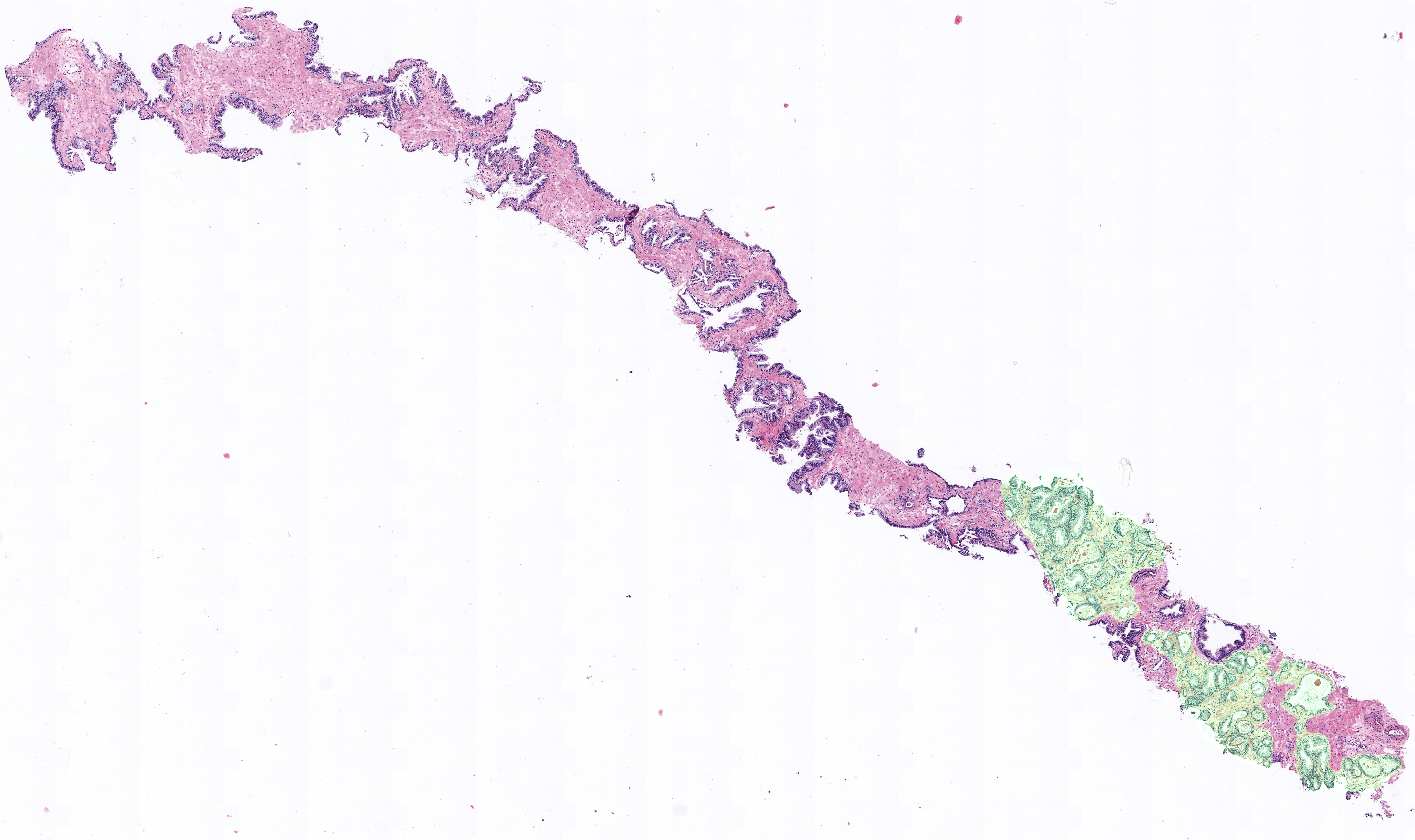}};
    \end{tikzpicture}}}
    \renewcommand{\thesubfigure}{c}
    \subfloat[\label{region2c}]{\resizebox{0.40\textwidth}{!}{
    \begin{tikzpicture}[spy using outlines={circle,yellow,magnification=10,size=3cm, connect spies}]
    \node {\pgfimage[interpolate=true,height=5cm]{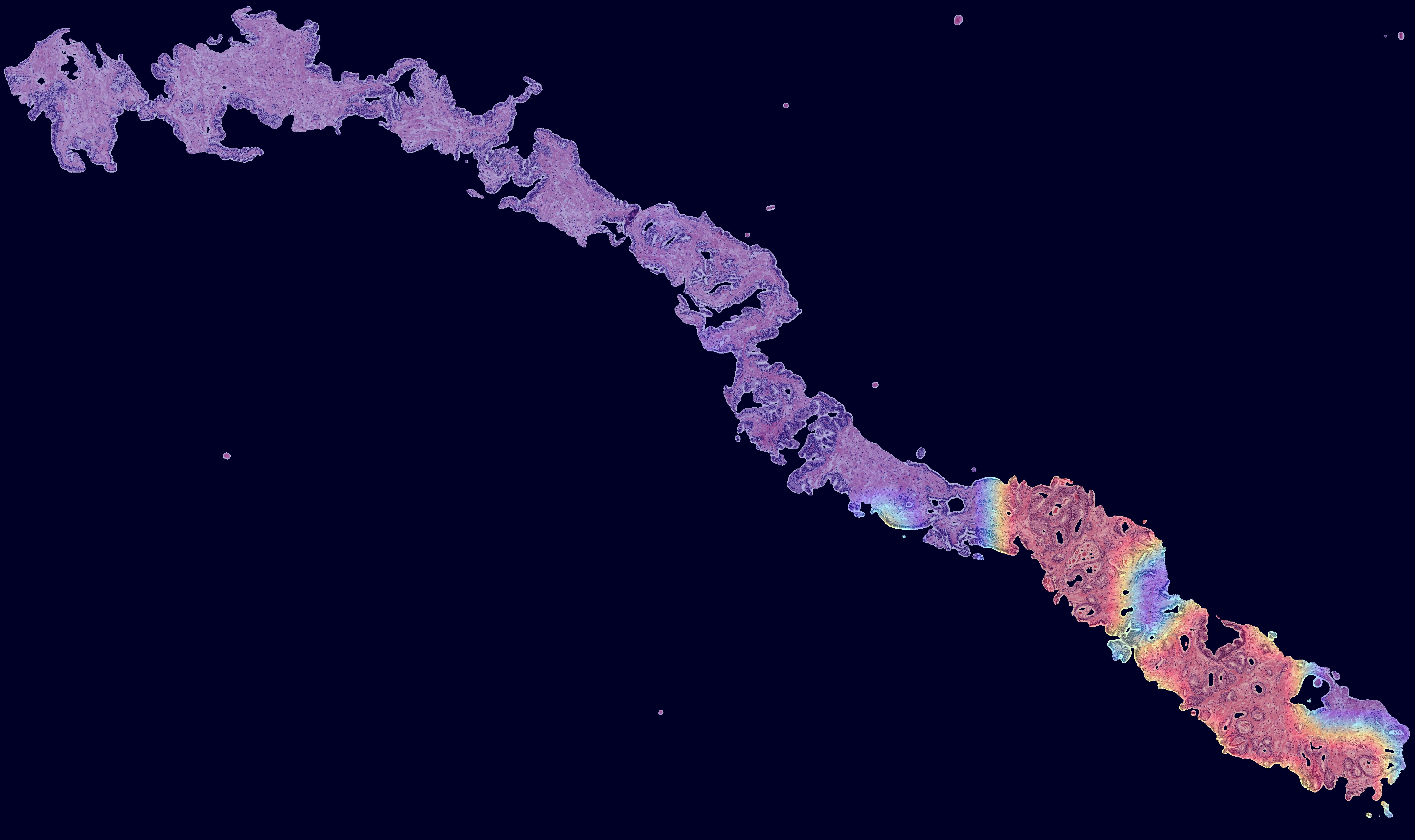}};
    \spy on (2,-0.6) in node [left] at (-1,-1);
    \spy on (3.3,-2) in node [left] at (4,1.5);
    \end{tikzpicture}}}
    
    \renewcommand{\thesubfigure}{b}
    \subfloat[\label{region2b}]{\resizebox{0.40\textwidth}{!}{
    \begin{tikzpicture}[spy using outlines={circle,yellow,magnification=2.5,size=3cm, connect spies}]
    \node {\pgfimage[interpolate=true,height=5cm]{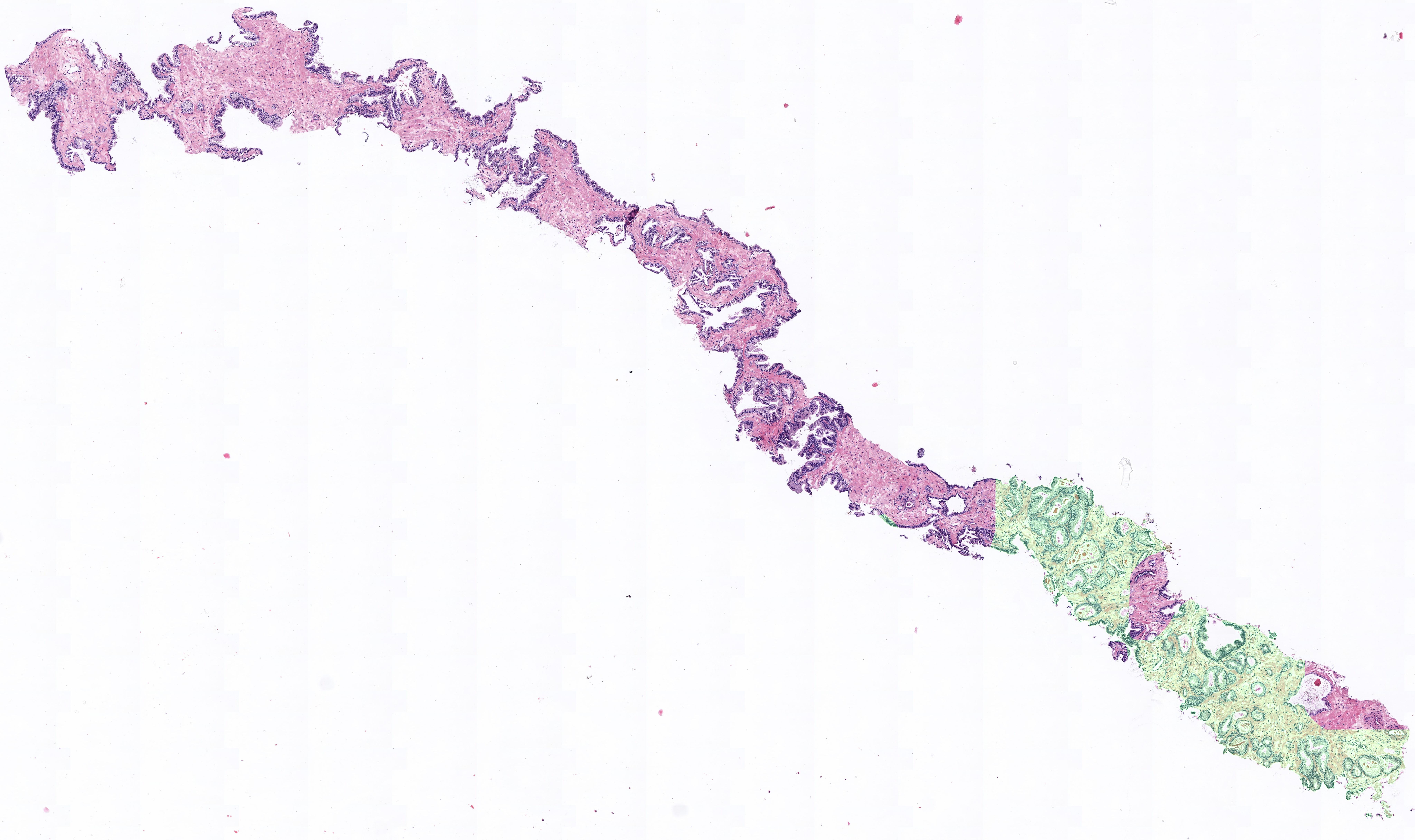}};
    \spy on (0.5,0) in node [left] at (-1,-1);
    \spy on (-0.1,0.7) in node [left] at (4,1.5);
    \end{tikzpicture}}}
    \renewcommand{\thesubfigure}{d}
    \subfloat[\label{region2d}]{\resizebox{0.40\textwidth}{!}{
    \begin{tikzpicture}
    \node {\pgfimage[interpolate=true,height=5cm]{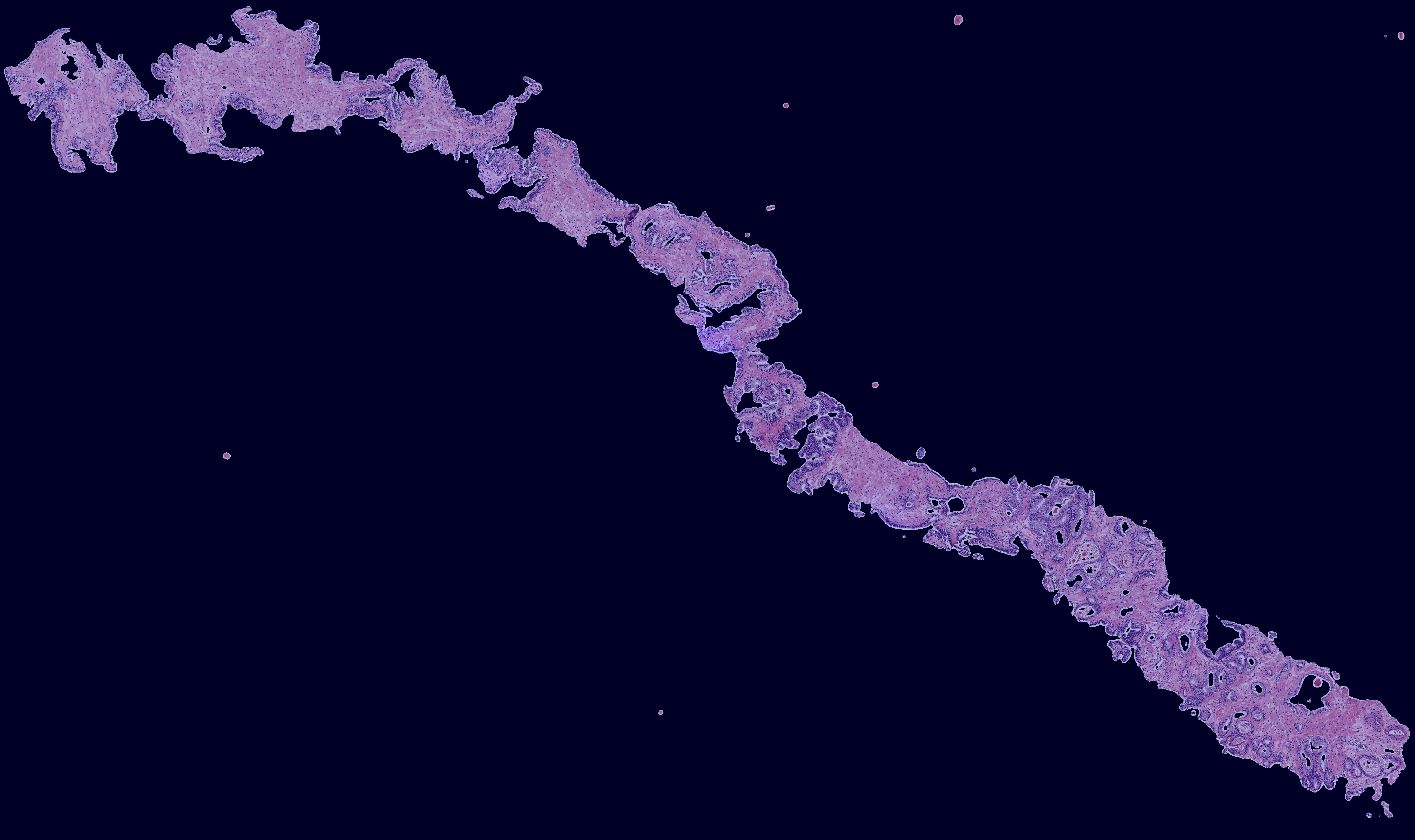}};
    \end{tikzpicture}}}

    \subfloat{\resizebox{0.40\textwidth}{!}{
    }}
    \renewcommand{\thesubfigure}{e}
    \subfloat[\label{region2e}]{\resizebox{0.40\textwidth}{!}{
    \begin{tikzpicture}
    \node {\pgfimage[interpolate=true,height=5cm]{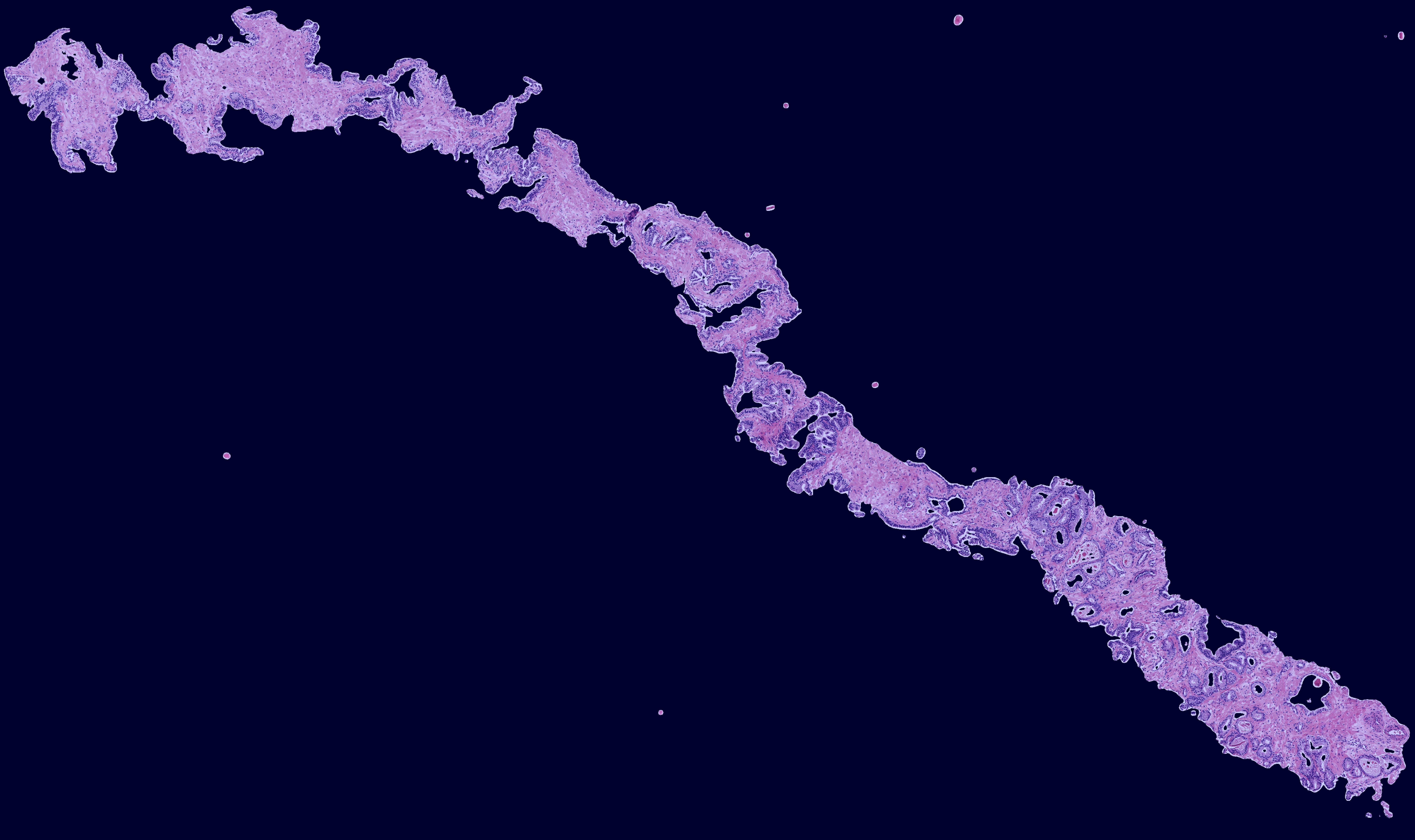}};
    \end{tikzpicture}}}

    \caption{Whole slide image level prediction of a biopsy diagnosed as Gleason Score $3+3=6$. (a): manual annotations, (b): system predictions. Green:  GG3, Blue:  GG4, red: GG5. (c): GG3 heatmap, (d): GG4 heatmap, (e): GG5 heatmap.}
    \label{fig:region2}
\end{figure}

\begin{figure}
    \centering
    
    \renewcommand{\thesubfigure}{a}
    \subfloat[\label{region3a}]{\resizebox{0.40\textwidth}{!}{
            \begin{tikzpicture}
            \node {\pgfimage[interpolate=true,height=5cm]{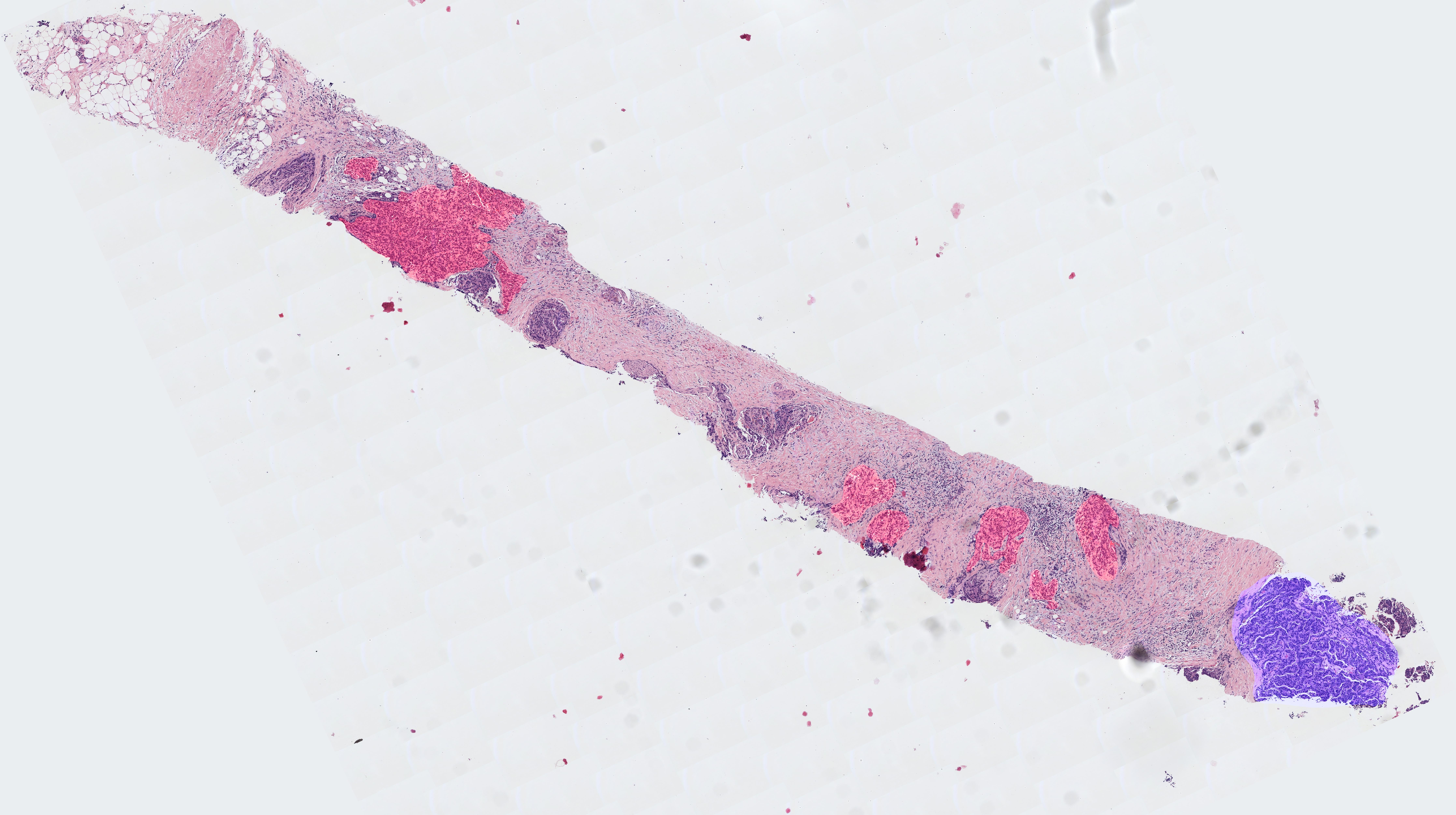}};
    \end{tikzpicture}}}
    \renewcommand{\thesubfigure}{c}
    \subfloat[\label{region3c}]{\resizebox{0.40\textwidth}{!}{
            \begin{tikzpicture}
            \node {\pgfimage[interpolate=true,height=5cm]{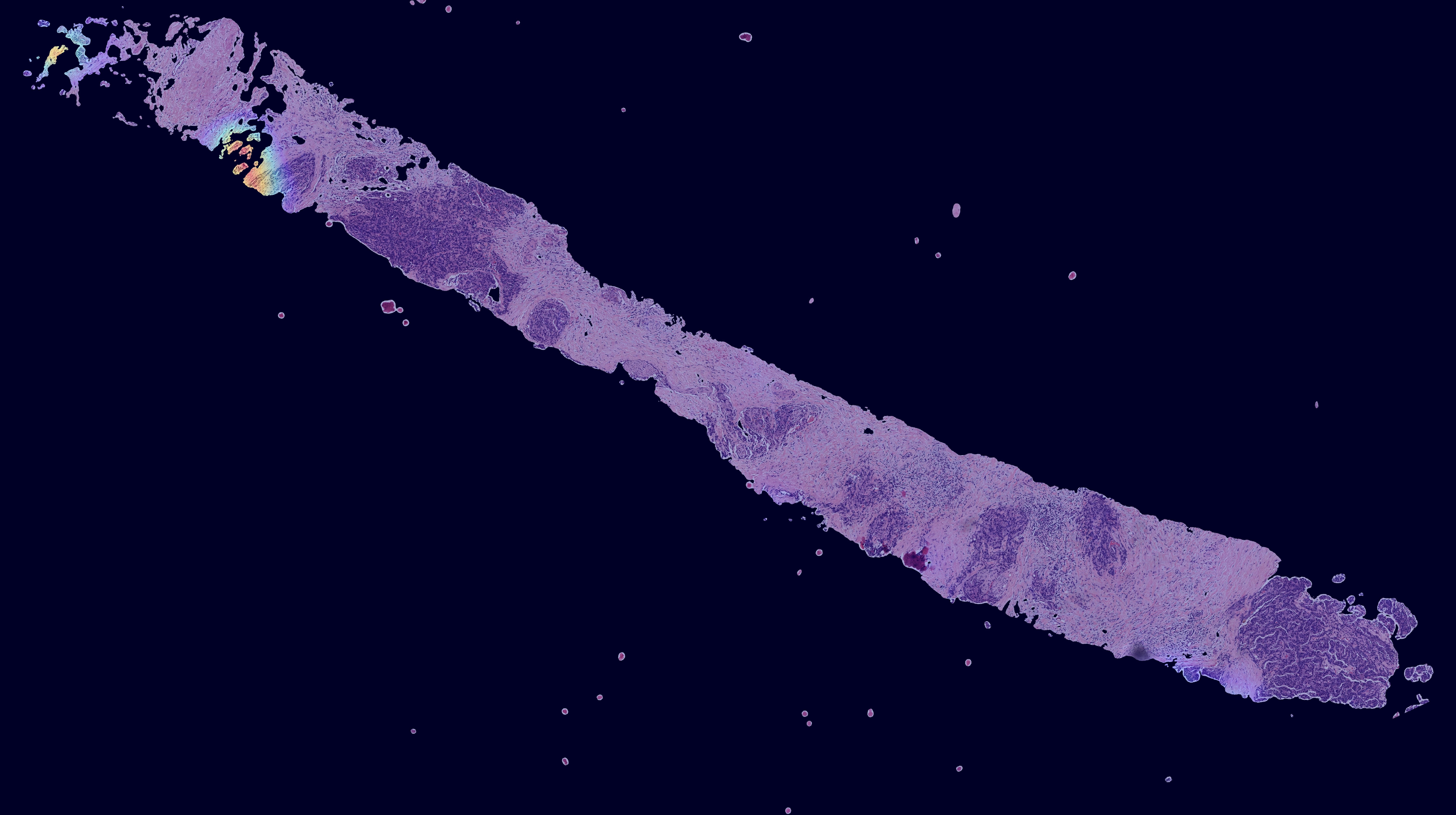}};
    \end{tikzpicture}}}
    
    \renewcommand{\thesubfigure}{b}
    \subfloat[\label{region3b}]{\resizebox{0.40\textwidth}{!}{
            \begin{tikzpicture}[spy using outlines={circle,yellow,magnification=15,size=3cm, connect spies}]
            \node {\pgfimage[interpolate=true,height=5cm]{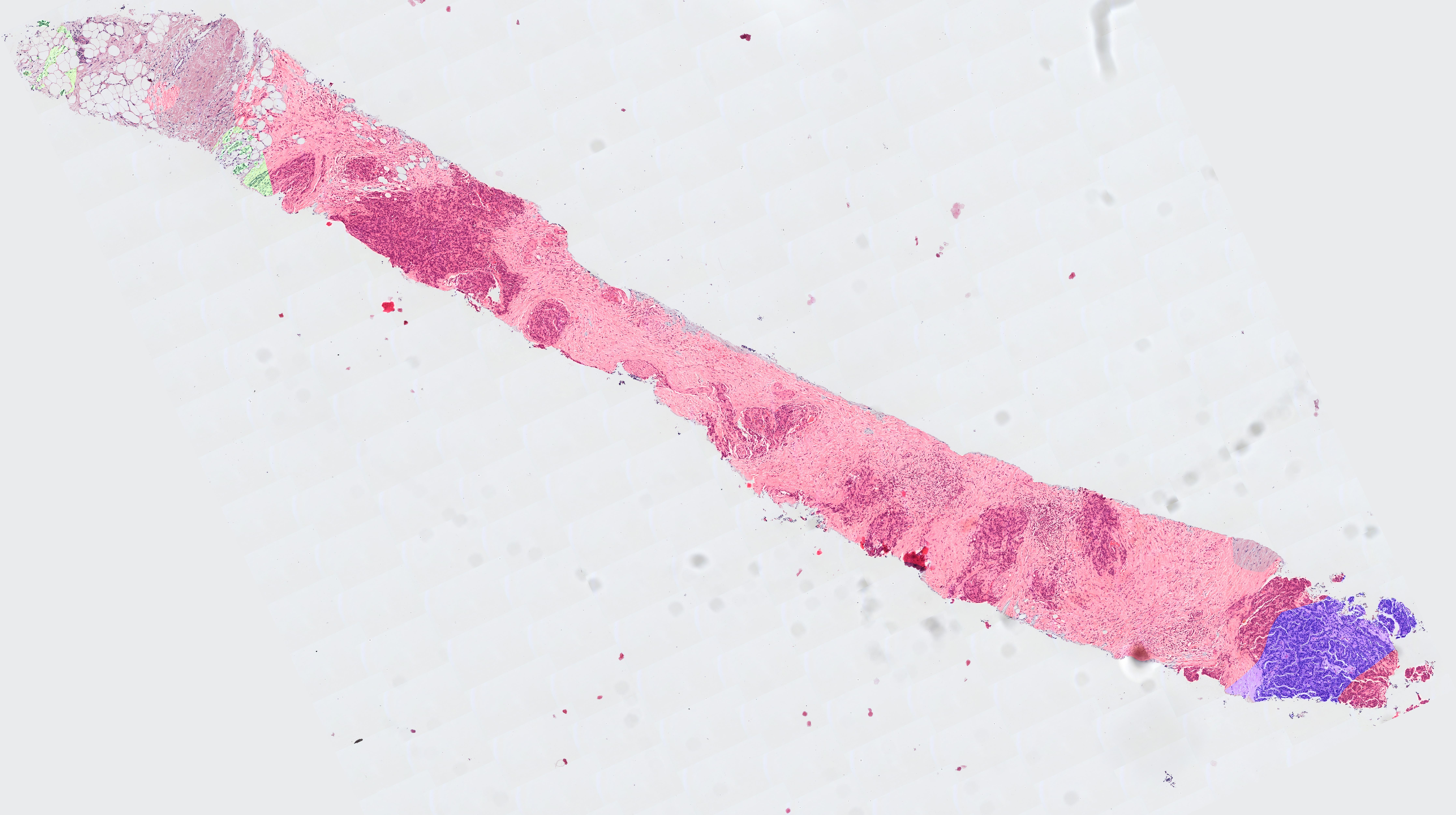}};
            \spy on (0.2,-0.2) in node [left] at (-1,-1.5);
            \spy on (-0.7,0.7) in node [left] at (4,1.5);
    \end{tikzpicture}}}
    \renewcommand{\thesubfigure}{d}
    \subfloat[\label{region3d}]{\resizebox{0.40\textwidth}{!}{
            \begin{tikzpicture}[spy using outlines={circle,yellow,magnification=10,size=3cm, connect spies}]
            \node {\pgfimage[interpolate=true,height=5cm]{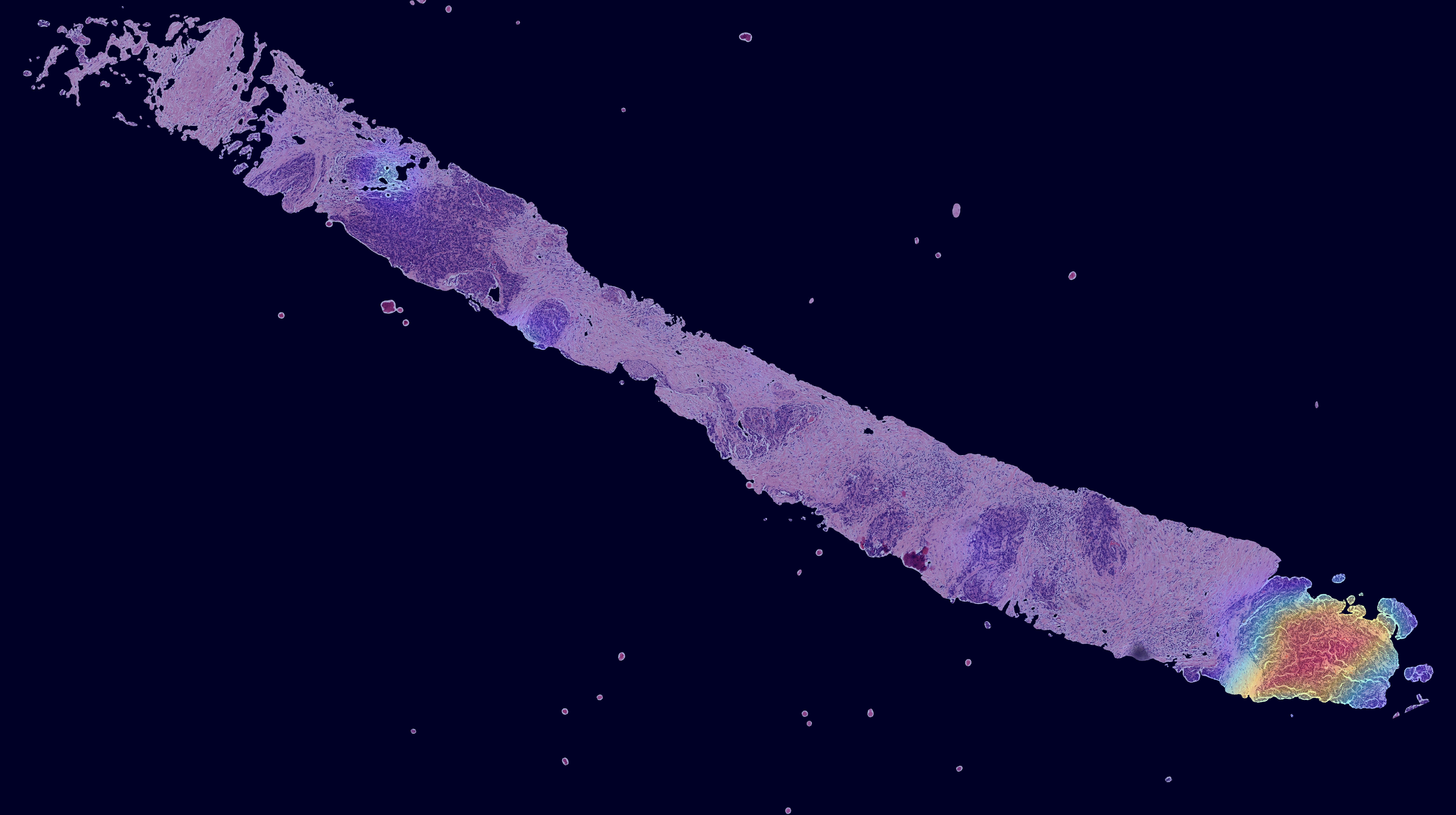}};
            \spy on (3.5,-1.5) in node [left] at (4,1.5);
    \end{tikzpicture}}}

    \subfloat{\resizebox{0.40\textwidth}{!}{
    }}
    \renewcommand{\thesubfigure}{e}
    \subfloat[\label{region3e}]{\resizebox{0.40\textwidth}{!}{
            \begin{tikzpicture}[spy using outlines={circle,yellow,magnification=10,size=3cm, connect spies}]
            \node {\pgfimage[interpolate=true,height=5cm]{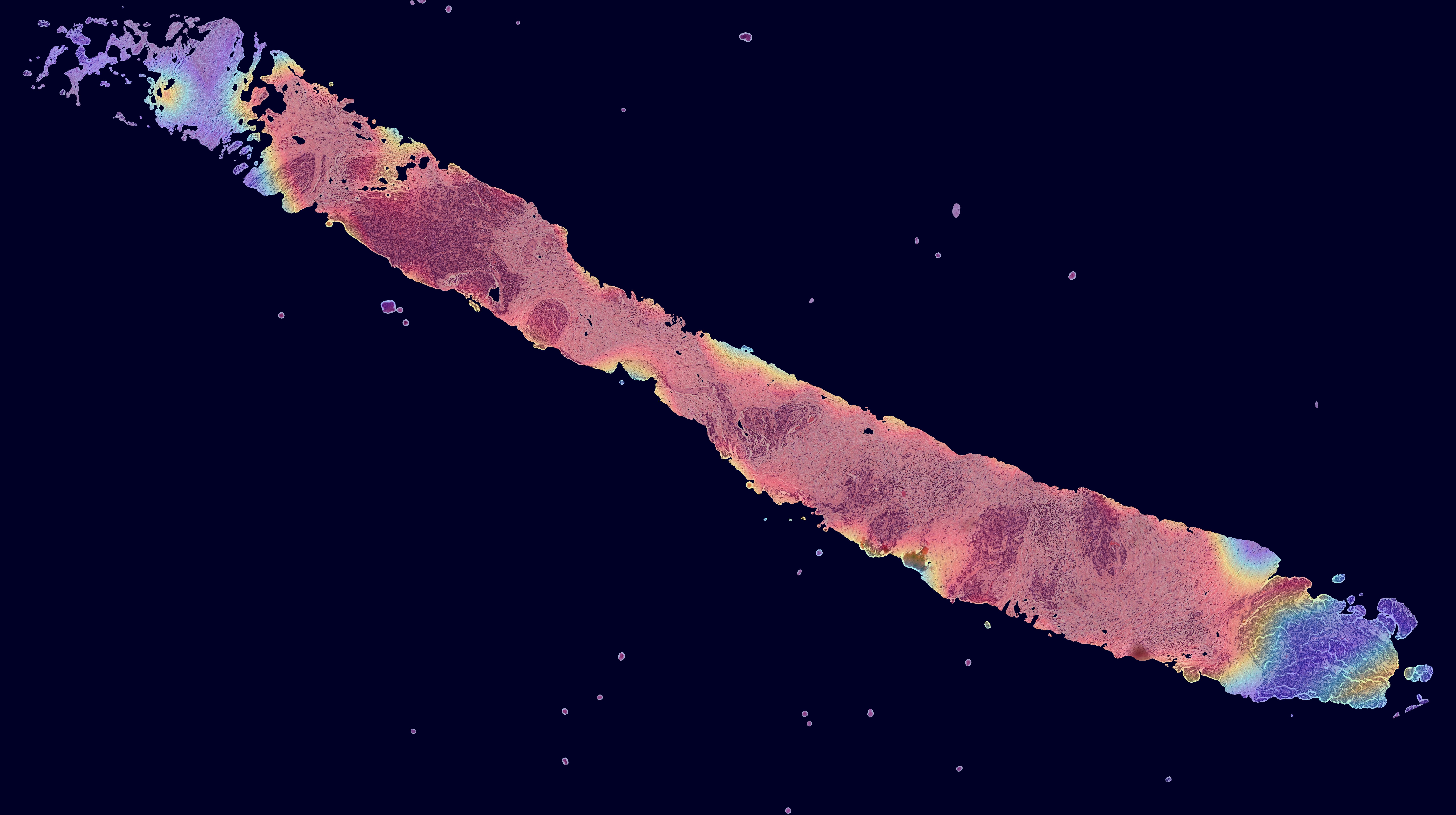}};
            \spy on (-1.75,1) in node [left] at (-1,-1.5);
            \spy on (1.7,-0.9) in node [left] at (4,1.5);
    \end{tikzpicture}}}

    \caption{Whole slide image level prediction of a biopsy diagnosed as Gleason Score $5+5=10$ (a): manual annotations, (b): system predictions. Green:  GG3, Blue:  GG4, red: GG5. (c): GG3 heatmap, (d): GG4 heatmap, (e): GG5 heatmap.}
    \label{fig:region3}
\end{figure}

\begin{figure}[htb]
    \centering
      \subfloat{\includegraphics[width=1\linewidth]{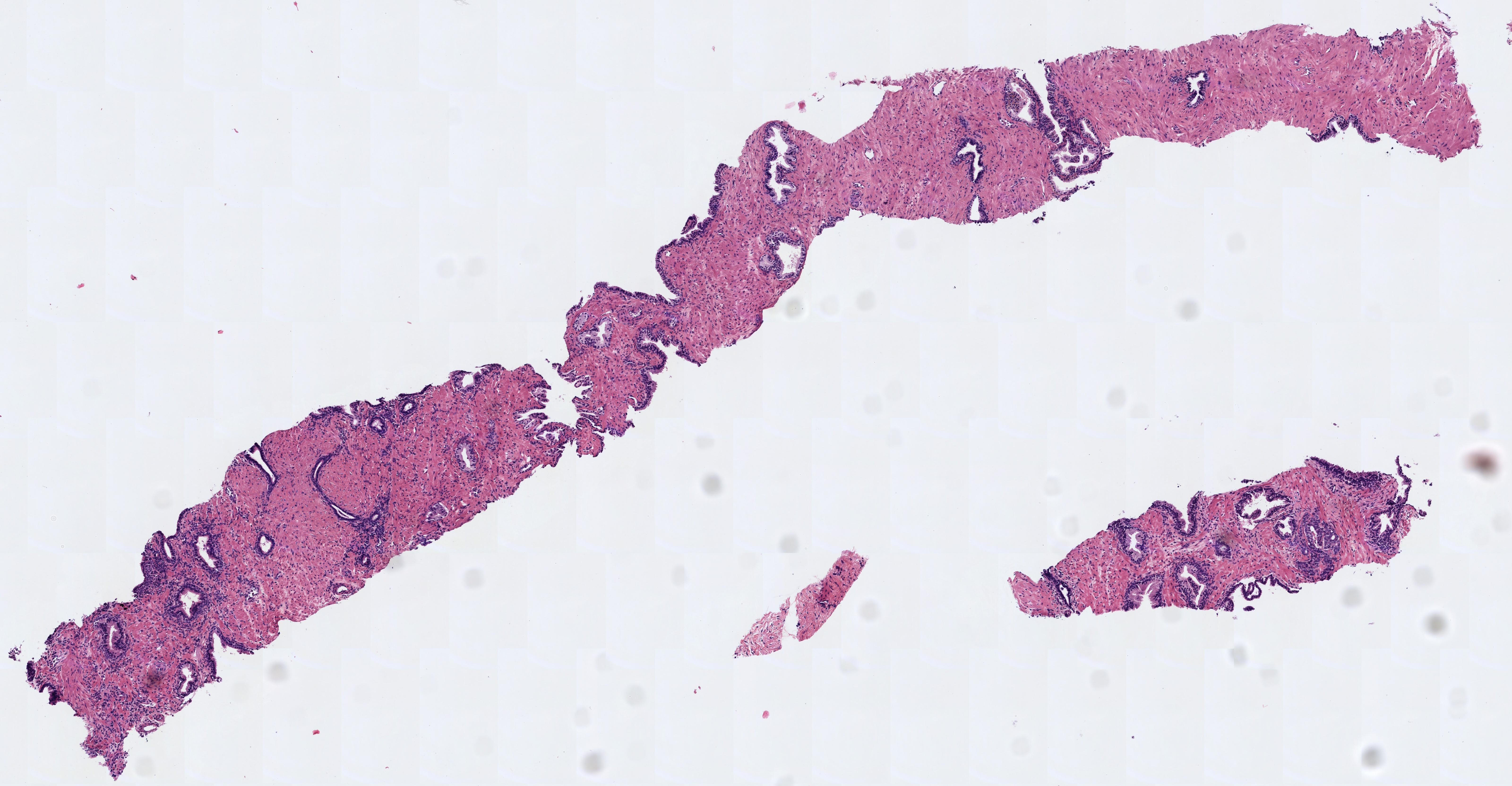}}
    \caption{Non-cancerous biopsy without Gleason grades detected by the model.}
    \label{fig:region4}
\end{figure}

Different examples of the test subset are presented in Figures \ref{fig:region1}, \ref{fig:region2}, and \ref{fig:region3}. These figures are organised as follows: in the first column, the WSI with pixel-level annotations (a) and pixel-level predictions (b) are presented, while in the second, the heatmaps of GG3 (c), GG4 (d) and GG5 (e) are shown from top to bottom, respectively. The regions of interest in the WSIs are highlighted with a higher resolution window to facilitate visualisation. The example in Figure \ref{fig:region1} is a biopsy with Gleason score $3+4=7$, the biopsy in Figure \ref{fig:region2} corresponds to a $3+3=6$ sample and the case in Figure \ref{fig:region3}, $5+5=10$. Finally, a non-cancerous case is presented in Figure \ref{fig:region4}.

In the case presented with Gleason score $3+4=7$ (see Figure \ref{fig:region1}), the GG3 and GG4 regions are correctly classified. In a subsequent review of this case, pathologists detected that some glands in the right region without pathologist's annotations in the ground truth and classified as GG3 by the model were actually cancerous patterns. Additionally, the few non-cancerous dilated and fusiform glands were correctly classified as non-cancerous (see Figure \ref{fig:region1} (b), regions of interest highlighted). Regarding the biopsy with Gleason score of $3+4=7$, the model correctly detects the region with GG3 glands, but due to the patch resolution ($512^{2}$ pixels) some nearby stroma regions are highlighted as cancerous. Finally, analysing the case with a score of $5+5=10$, a papilar GG4 pattern is being correctly detected. The same occurs in the GG5 regions with isolated cells and pseudorosetting patterns. Nevertheless, in regions with a score of $GS\geq9$ some stroma regions are frequently highlighted as GG5 by the model. This phenomenon does not occur in stroma of biopsies with $GS<9$, as can be seen in the other cases. This fact suggests that the model could be detecting some hidden pattern of interest in the structure of the stroma in these regions.

Then, the percentages corresponding to each grade per WSI were obtained as specified in the methodology (Section \ref{chap:WSIscoring}). The proposed architecture $MLP$ was then trained using as input the percentages obtained in the cross-validation subset. Adam optimiser was used, with a learning rate of $0.01$, and a constant decay to zero over the $2000$ epochs. The batch size was $32$. The training strategy was leave-one-out. 

\begin{figure}[htb]
    \centering
      \subfloat[\label{cmScoringa}]{\includegraphics[width=.48\linewidth]{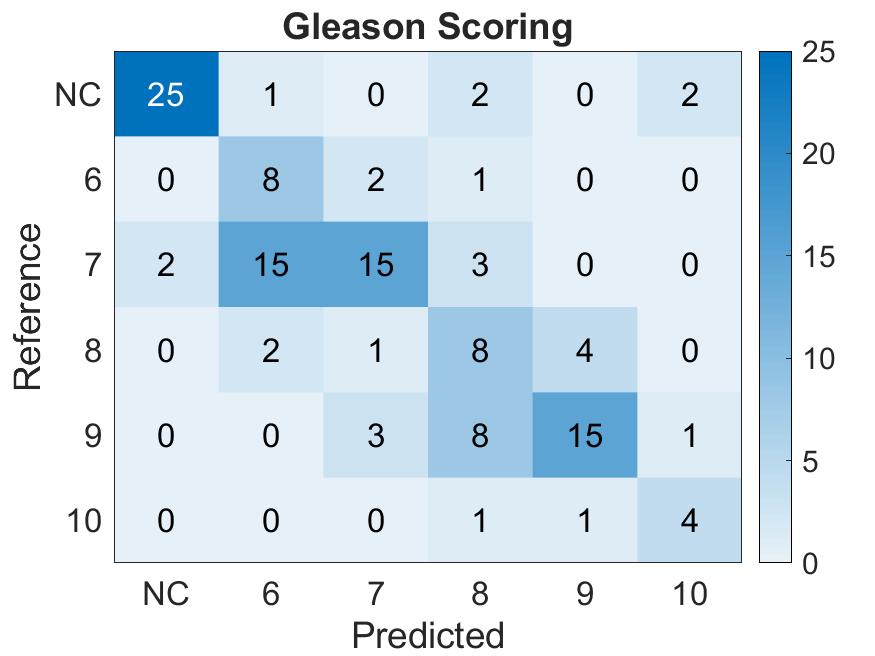}}
      \hspace*{\fill}
      \subfloat[\label{cmScoringb}]{\includegraphics[width=.48\linewidth]{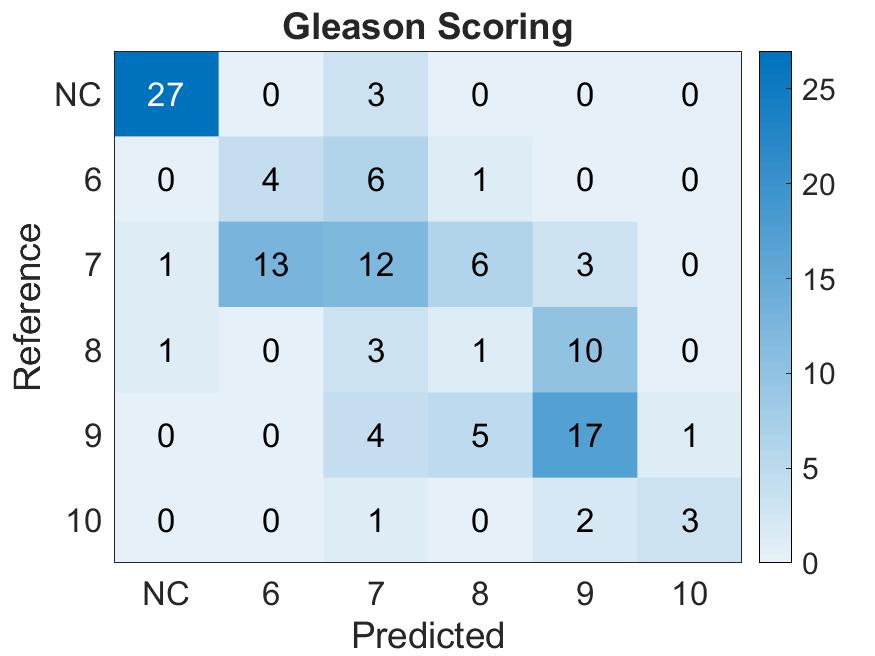}}
      \hspace*{\fill}
    \caption{Confusion Matrix of the whole slide image level Gleason scoring in the validation cohorts. (a): Method proposed in \cite{Arvaniti2018AutomatedLearning}; (b): $MLP$ model. }
    \label{fig:scoringCM}
\end{figure}

This proposed approach is compared with the method proposed by Arvaniti \cite{Arvaniti2018AutomatedLearning} using $T=10\%$ as minimum number of pixels with a certain label to be consider the corresponding grade in the WSI grading. The confusion matrix at biopsy level obtained for both methods is presented in Figure \ref{fig:scoringCM}, and Cohen's quadratic kappa ($\kappa$) was calculated as a figure of merit.

The $\kappa$ value obtained for Arvaniti's approach was $0.7693$, in line with the results presented in  \cite{Arvaniti2018AutomatedLearning} using their own database (using TMAs), where the obtained $\kappa$ value was $0.75$. Better results were obtained with the proposed model $MLP$ (see Figure \ref{fig:scoringCM} (b)), obtaining a $\kappa$ value of $0.8177$. The main difference between methods was observed in few samples misclassified as Gleason score $8$ and Gleason score $10$ by Arvaniti's proposal which were correctly classified by our model. Therefore, our proposed strategy seems to model better the pathologist's decision to assign a Gleason score to the full image of the slide than the previous scoring methodology. The results obtained in the test subset by $MLP$ model are similar to those obtained for the validation cohorts, with a $\kappa$ value $0.8168$.


\section{Conclusions and future work}
\label{conclusions}

In this work, we have proposed and validated end-to-end approaches to automatically support the pathologists analysis of prostate whole slide images. This support includes the pixel-level prediction of Gleason grades, cribriform patterns detection, calculation of the percentage of each grade in the tissue and finally the scoring of the entire biopsy.

We have compared fine-tuned state-of-the-art architectures and  self-designed convolutional neural network architectures trained from scratch for the patch-level Gleason grades prediction. In addition, we have discussed the use of a global-max-pooling and global-average-pooling layers in the top model for this application. The use of global-max pooling has showed interesting properties in the model trained from scratch. It supports the use of shallow architectures with a small receptive field and a reduced amount of parameters, diminishing one of the main drawbacks of training from scratch: the over fitting to the training set. Thus, with a concise model composed of three convolutional layers, we have achieved the best results in our data set, reaching a Cohen`s quadratic kappa of $0.77$ in the test images. Furthermore, by just re-training the filter weights of the last convolutional layer, we have predicted the presence of cribriform regions in patches with Gleason grade $4$, with an AUC value of $0.82$ in the test subset. To the best of the authors's knowledge, this is the first work contemplating the automatic detection of cribriform patterns in prostate histology images. We also have studied the interpretability of the developed deep-learning models by means of Class Activation Maps. Additionally, we have obtained probability heat maps indicating the presence of the different Gleason grades in the whole slide image. Finally, making use of the percentage of non-cancerous, Gleason grade $3$, $4$, and $5$ tissues in the biopsy we have predicted its combined Gleason score through a multi-layer perceptron, reaching a Cohen's quadratic kappa of $0.8168$ in the test cohort. This model reproduces better the decision-making of the pathologist reporting the biopsy score than previous ones based on just assigning the two first grades with a higher percentage.

The limitations of the study naturally include the intra-observer variability of the annotator. This fact is not present on the trained algorithm, but it could affect the figures of merit obtained. Additionally, the large heterogeneity inside each Gleason grade makes difficult to balance the different folds, representing all the different patterns of the Gleason grades in all the training and testing groups. 


It is important to note that this work brings an important contribution to the scientific community: the SICAPv2 database, the largest public database containing pixel-level annotations of prostate biopsies.

Further research will focus on developing convolutional-neural-network architectures that combine low and high-level features in the classification stage, as well as the inclusion in those models the prediction of all the individual cancerous patterns (i.e. ill-fused, papillary or large-fused) as the cribriform one, in an end-to-end training. Furthermore, the SICAPv2 database will be enlarged with additional annotated whole slide images.

\section*{References}

\bibliography{mybibfile,references}

\end{document}